\documentclass[12pt]{article}
\usepackage{amsmath,amssymb,bm,epsfig,color,graphicx,cite,braket,mathrsfs,slashed,multirow}
\usepackage{tikz-feynhand}
\renewcommand{\thefootnote}{\fnsymbol{footnote}}

\begin{document}

\title{
\begin{flushright}
\begin{minipage}{0.21\linewidth}
\normalsize
CTPU-PTC-21-32 \\*[50pt]
\end{minipage}
\end{flushright}
{\Large \bf 
Unstable Cosmic Neutrino Capture
\\*[20pt]}}

\author{
Kensuke Akita$^{1,2}$\footnote{
E-mail address: kensuke8a1@ibs.re.kr},\
Gaetano Lambiase$^{3,4}$\footnote{
E-mail address: lambiase@sa.infn.it}
  and\
Masahide Yamaguchi$^{1}$\footnote{
E-mail address: gucci@phys.titech.ac.jp}\\*[20pt]
$^1${\it \small
Department of Physics, Tokyo Institute of Technology,
Tokyo 152-8551, Japan} \\
$^2${\it \small
Center for Theoretical Physics of the Universe, Institute for Basic Science,
Daejeon 34126, Korea} \\
$^3${\it \small
INFN Sezione di Napoli, Gruppo collegato di Salerno, I-84084 Fisciano (SA), Italy} \\
$^4${\it \small
Dipartimento di Fisica "E.R. Caianiello", Universit\`a di Salerno, I-84084 Fisciano (SA), Italy} \\*[50pt]
}

\date{
\centerline{\small \bf Abstract}
\begin{minipage}{0.9\linewidth}
\medskip \medskip \small 
Future direct observations of the Cosmic Neutrino Background (C$\nu$B) have the potential to explore a neutrino lifetime, especially in the region of the age of the universe, $t_0=4.35\times 10^{17}\ {\rm s}$.
We forecast constraints on neutrino decay via capture of the C$\nu$B on tritium, with emphasis on the PTOLEMY-type experiment.
In addition, in some cases of invisible neutrino decay into lighter neutrinos in the Standard Model and invisible particles, 
we can constrain not only the neutrino lifetime but also the masses of the invisible particles.
For this purpose, we also formulate the energy spectra of the lighter neutrinos produced by 2-body and 3-body decays, and those of the electrons emitted in the process of the detection of the lighter neutrinos.
\end{minipage}
}

\maketitle{}
\clearpage
\tableofcontents
\clearpage

\renewcommand{\thefootnote}{\arabic{footnote}}
\setcounter{footnote}{0}
%\vspace{35pt}

\section{Introduction}
\label{sec:1}

Neutrino oscillation experiments have established that neutrinos have tiny masses and flavor mixing exists in the neutrino sector.
However, the mechanism that generates the tiny neutrino masses remains one of the most important mysteries in particle physics.
Thus, neutrino oscillations imply the existence of physics beyond the Standard Model (SM) and non-standard interactions in the neutrino sector.
In addition, these experiments bring us information only on the differences of the neutrino mass squared, that is $\Delta m^2_{ij}=m_{\nu_i}^2-m_{\nu_j}^2$ \cite{Esteban:2020cvm},
\begin{align}
\Delta m_{21}^2 \simeq (8.6\ {\rm meV})^2\ \ \ \ {\rm and}\ \ \ \ |\Delta m_{3l}^2|  \simeq (50\ {\rm meV})^2,
\label{SD}
\end{align}
where $l=1,2$. The neutrino mass ordering and the absolute values of neutrino masses (in particular, the lightest neutrino mass $m_{\rm lightest}$) have not yet been known. 
The mass ordering has two possibilities, $m_{\nu_3}>m_{\nu_2}>m_{\nu_1}$ called Normal Ordering (NO) and $m_{\nu_2}>m_{\nu_1}>m_{\nu_3}$ called Inverted Ordering (IO).
The currently best laboratory limit on the neutrino mass itself is $m_{\bar{\nu}_e}\equiv\sqrt{\sum_i|U_{ei}|^2m_{\nu_i}^2}<0.8$ eV at 90$\%$ CL that is reported by the KATRIN experiment\cite{Aker:2021gma}. 
In the future, the planned measurement of 1000 days will allow us to reach $m_{\bar{\nu}_e}\lesssim 0.2$ eV at 90$\%$ CL.
When the neutrino masses are degenerate as $m_{\nu_i}\simeq m_{\rm lightest}$, the above limit could directly constrain $m_{\rm lightest}$ thanks to $m_{\bar{\nu}_e}\simeq m_{\rm lightest}$.

One of the significant phenomena induced by massive neutrinos is neutrino decay. With the SM interactions, heavier neutrinos can radiatively decay into lighter neutrinos and photons through a non-zero magnetic moment induced by one-loop interactions \cite{Fujikawa:1980yx}.
Due to the suppression of one-loop interactions, the neutrino lifetime is much larger than the age of the universe $\tau_{\nu}\gtrsim10^{36}(m_{\nu}/{\rm eV})^{-5}\ {\rm yr}$, where $m_{\nu}$ is the neutrino mass.\footnote{Then, we practically look on the neutrino lifetime as infinitely long for the massive neutrinos without non-standard decay modes.} 
On the other hand, non-standard interactions can induce invisible neutrino decays, where heavier neutrinos decay into lighter neutrinos and undiscovered light particles. 
With such a decay channel, the heavier neutrinos can have much shorter lifetimes.
One of the extensions that induce invisible neutrino decays is the majoron model \cite{Chikashige:1980ui, Gelmini:1980re, Schechter:1981cv}.
In this scenario, the neutrino mass is generated by the breaking of a global lepton number symmetry, whose pseudo-Nambu Goldstone boson is called majoron.

In radiative decay scenarios, a neutrino lifetime is severely constrained by electron-neutrino scattering experiments as $\tau_{\nu}\gtrsim10^{18}\ {\rm yr}$ \cite{Beda2013g, Borexino:2017fbd}. The lower bound from astrophysical observations \cite{Raffelt:1990pj, Viaux:2013hca, Arceo-Diaz:2015pva} and cosmological 21 cm  absorption signals \cite{Chianese:2018luo} is $\tau_{\nu}\gtrsim10^{20}\ {\rm yr}$. 
Cosmic Microwave Background (CMB) spectral distortions also constrain a neutrino lifetime as $\tau_{\nu}\gtrsim10^{12}\ {\rm yr}$ \cite{Mirizzi:2007jd, Aalberts:2018obr}.

However, for invisible neutrino decays, the lower bound of a neutrino lifetime is drastically weaker 
because it is hard to probe particles produced by invisible decays.
In addition, the decay rates of ultra-relativistic neutrinos, which are the currently observed neutrinos, are highly suppressed by the Lorentz factor.
Experimental constraints on invisible neutrino decays come mainly from solar and reactor neutrino experiments.
The strongest constraint on the lifetime of $\nu_2$ is $\tau_{\nu_2}/m_{\nu_2}\gtrsim1.5 \times 10^{-3}\ {\rm s\ eV}^{-1}$ \cite{Aharmim:2018fme}.
In addition, the lower limits on the lifetime of $\nu_1$ and $\nu_3$ are  $\tau_{\nu_1}/m_{\nu_1}\gtrsim4 \times 10^{-3}\ {\rm s\ eV}^{-1}$ \cite{Berryman:2014qha} and  $\tau_{\nu_3}/m_{\nu_3}\gtrsim2.2 \times 10^{-5}\ {\rm s\ eV}^{-1}$ \cite{Funcke:2019grs}, respectively.
Cosmological constraints on a neutrino lifetime through invisible decays also come from observations of the CMB anisotropies and Big Bang Nucleosynthesis (BBN). The measurements of the CMB anisotropies impose the strongest constraint of $\tau_{\nu}\gtrsim1.2\times 10^9\ {\rm s}\ (m_{\nu}/50\ {\rm meV})^3$ \cite{Archidiacono:2013dua, Escudero:2019gfk} when non-relativistic neutrinos decay, and $\tau_{\nu}\gtrsim4\times 10^5\ {\rm s}\ (m_{\nu}/50\ {\rm meV})^5$ \cite{Barenboim:2020vrr} when relativistic neutrinos decay. The observations of BBN also impose the constraints of $\tau_{\nu}\gtrsim10^{-3}\ {\rm s}$ \cite{Escudero:2019gfk}, and
the bound from Supernova 1987A observations is $\tau_{\nu}/m_{\nu}>10^{5}\ {\rm s}\ {\rm eV^{-1}}$ \cite{Kamiokande-II:1987idp}.
In the near future, the constraints on neutrino decays will be pursued more by measurements of atmospheric and reactor neutrinos \cite{Abrahao:2015rba, Coloma:2017zpg, Choubey:2017dyu, Choubey:2017eyg, Ascencio-Sosa:2018lbk, Tang:2018rer, deSalas:2018kri}, solar neutrinos \cite{Huang:2018nxj}, high energy astrophysical neutrinos \cite{Beacom:2002vi,Pagliaroli:2015rca, Bustamante:2016ciw, Denton:2018aml, Abdullahi:2020rge}, a galactic supernova \cite{Tomas:2001dh, Lindner:2001th, Ando:2004qe, deGouvea:2019goq} and the diffuse supernova neutrino background \cite{Ando:2003ie, Fogli:2004gy, DeGouvea:2020ang}.

Future direct detections of the Cosmic Neutrino Background (C$\nu$B) will place strong constraints on neutrino decays, in particular, much stronger limits on invisible neutrino decays since cosmic neutrinos are freely streaming almost during the age of the universe, $t_0=4.35 \times 10^{17}\ {\rm s}$ \cite{Aghanim:2018eyx}. 
In addition, since the average magnitude of momenta for neutrinos in the current universe $p_0$ is $\langle p_0 \rangle \sim 0.53\ {\rm meV}$, which is smaller than the observed two mass-squared differences, at least two mass-eigenstates of neutrinos are non-relativistic. Their decay rates in the current universe are not suppressed by the Lorentz factor. From these facts, we can impose the constraint on neutrino lifetimes of $\tau_{\nu}\sim t_0$ via future direct detections for the C$\nu$B. If cosmic neutrinos would have decayed until today, this result can place upper bounds of neutrino lifetimes,
complementing the current lower bounds of neutrino lifetimes. 

The most promising method\footnote{The various other methods for direct detections are also proposed\cite{Stodolsky:1974aq, Shvartsman:1982sn, Langacker:1982ih, Duda:2001hd, Domcke:2017aqj, Bauer:2021txb}.} of direct detections of the C$\nu$B is neutrino capture on $\beta$-decaying nuclei \cite{Weinberg:1962zza, Cocco:2007za}, in particular on tritium target \cite{Lazauskas:2007da, Blennow:2008fh, Long:2014zva, Roulet:2018fyh, Akita:2020jbo}, via the inverse $\beta$-decay process, $\nu_i+\mathrm{^3H}\rightarrow e^-+\mathrm{^3He}$ (see refs.\cite{Li:2010sn, Arteaga:2017zxg ,McKeen:2018xyz,Chacko:2018uke, Bondarenko:2020vta, Alvey:2021xmq} for studies of physics beyond the SM via cosmic neutrino capture on tritium). This method has at least two merits:
(i) this process includes no threshold energy, (ii) thanks to the energy injection of cosmic neutrinos, its energy of emitted electrons appears above the $\beta$-decay endpoint, $\mathrm{^3H}\rightarrow \bar{\nu}_i+ e^-+\mathrm{^3He}$. The challenges of the cosmic neutrino capture on tritium lie in large tritium mass and high resolution of emitted electron energy to distinguish its signature and the noise from the $\beta$-decay background (If tritium is solid state, an additional limitation for the energy resolution might be occur due to the fundamental uncertainty principle \cite{Cheipesh:2021fmg,Nussinov:2021zrj}). Recently, as a cosmic neutrino capture experiment, the PTOLEMY-type experiment has been proposed, in which 100 grams of tritium target and the graphene-based detector will be used\cite{Betti:2018bjv, Betti:2019ouf}.

In this work, we forecast constraints on neutrino decays via cosmic neutrino capture on tritium, with emphasis on the PTOLEMY-type experiment. In particular, we consider the invisible decays of heavier SM neutrinos to lighter SM neutrinos in the current universe.
If such decays would be significant, the current energy distributions of the neutrinos will be different from those without such decays, for example, those in the SM.
Thus, the energy distributions of the neutrinos will give us a wealth of information. We formulate the energy distributions for cosmic neutrinos in 2-body and 3-body decays.
Then we forecast the expected signals in cosmic neutrino capture on tritium. 
In addition, we discuss future constraints not only on a neutrino lifetime but also on the mass of an invisible particle from the PTOLEMY-type experiment because the expected energy distributions would depend on the kinematics of the invisible neutrino decays.

This paper is organized as follows.
In the next section, we review the properties of the decays of the C$\nu$B.
In section~\ref{Sec3}, we discuss invisible decays of cosmic neutrinos induced by non-standard interactions. In particular, we estimate the relations between the suppression factors by invisible neutrino decays and the coupling constants of non-standard neutrino interactions. 
In section~\ref{Sec4}, we formulate lighter neutrino spectra produced by the decays of heavier neutrinos.
In section~\ref{Sec5}, the would-be observed spectra for unstable neutrinos and the future constraints on a neutrino lifetime and an invisible particle mass are given. Finally, section~\ref{Sec6} is devoted to conclusions. In the appendix, we show invisible neutrino decay rates in several cases.

%%%%%%%%%%%%%%%%%%%%%%%%%%%%%%%%%%%%%%%%%%%%%%%%%%%%%%%%%%%%%%%%%%%%%%%%%%%%%%%%%%%%%%%%%%%%%%%%%%
%%%%%%%%%%%%%%%%%%%%%%%%%%%%%%%%%%%%%%%%%%%%%%%%%%%%%%%%%%%%%%%%%%%%%%%%%%%%%%%%%%%%%%%%%%%%%%%%%%
%%%%%%%%%%%%%%%%%%%%%%%%%%%%%%%%%%%%%%%%%%%%%%%%%%%%%%%%%%%%%%%%%%%%%%%%%%%%%%%%%%%%%%%%%%%%%%%%%%
%%%%%%%%%%%%%%%%%%%%%%%%%%%%%%%%%%%%%%%%%%%%%%%%%%%%%%%%%%%%%%%%%%%%%%%%%%%%%%%%%%%%%%%%%%%%%%%%%%

\section{Cosmic neutrino decays}
\label{Sec2}

In this section, we review the decays of cosmic neutrinos. 
In particular, we consider the number density of decaying neutrinos in the current universe.
%This calculation can be applied to both radiative neutrino decays and invisible neutrino decays.

If cosmic neutrinos would decay until today, the neutrino number densities will be significantly suppressed depending on their lifetimes.
In such a scenario, the present neutrino number density per one degree of freedom is given by
\begin{align}
n_{\nu}(t_0)=e^{-\lambda_{\nu}} f_c n_{\nu}^0,
\end{align}
where $f_c$ is the enhancement factor for the gravitational clustering by our Galaxy and nearby galaxies\cite{Alvey:2021xmq, Singh:2002de, Ringwald:2004np, deSalas:2017wtt, Zhang:2017ljh, Mertsch:2019qjv}, and $e^{-\lambda_{\nu}}$ is the suppression factor for the possible neutrino decays. $n_{\nu}^0 \equiv n_{\nu}(t_d) [a(t_d)/a(t_0)]^3 \simeq 56\ {\rm cm^{-3}}$ is the would-be current number density of neutrinos per one degree of freedom in the SM (practically without decays), $t_0$ is the present cosmic time, $t_d$ is the decoupling time of neutrinos, $a(t)$ is the scale factor of the universe at the time $t$, and $n_{\nu}(t)$ is the neutrino number density per one degree of freedom at the time $t$. The enhancement factor $f_c$ depends on the mass of neutrinos and we display several values of $f_c$ in Table~\ref{tb:CND} for reference \cite{Mertsch:2019qjv}. 
Here, in terms of the time $t$ and the redshift $z$, $\lambda_{\nu}$ is given by \cite{Baerwald:2012kc, Long:2014zva}
\begin{align}
\lambda_{\nu}=\int^{t_0}_{t_d} \frac{dt}{\tau_{\nu}'}=\int^{z_{\rm d}}_0\frac{dz}{(1+z)H(z)\gamma(z)\tau_{\nu}},
\label{lambda}
\end{align}
where $\tau_{\nu}$ is the lifetime in the rest frame of neutrinos, boosted into the lifetime in the frame of the observer at epoch $z$, $\tau_{\nu}'(z)=\gamma(z)\tau_{\nu}$, by the Lorentz factor,
 \begin{align}
 \gamma(z)=\frac{E_{\nu}(z)}{m_{\nu}}=\sqrt{\frac{p_0^2}{m_{\nu}^2}(1+z)^2+1},
 \end{align}
 where $p_0$ is the momentum of a neutrino at $z=0$.
 $z_{\rm d}\simeq 6\times 10^9$ is the redshift of neutrino decoupling
 and the Hubble parameter $H(z)$ is given by
 \begin{align}
 H(z)=H_0\sqrt{\Omega_r(1+z)^4+\Omega_m(1+z)^3+\Omega_\Lambda},
 \end{align}
where $H_0\simeq 67.36\ {\rm km\ s^{-1}\ Mpc^{-1}}$ is the present Hubble parameter. $\Omega_r<10^{-4}$, $\Omega_m=0.3158$ and $\Omega_\Lambda=0.6842$ are the present (normalized) energy densities of radiation, matter, and cosmological constant, respectively \cite{Aghanim:2018eyx}.

The era close to the current epoch, $z\ll1$, contributes to the integral of Eq.~(\ref{lambda}) most efficiently since the integrand is at least suppressed by $(1+z)$.
In addition, if neutrinos are non-relativistic in the current universe, then the Lorentz factor at $z\ll1$ is $\gamma(z\ll1)\simeq 1$.
Thus, the expression of Eq.~(\ref{lambda}) for (currently) non-relativistic neutrinos is simplified as
\begin{align}
\lambda_{\nu}\simeq t_0/\tau_{\nu},
\label{applambda}
\end{align}
where $t_0$ is the current age of the universe. 
We have numerically confirmed that this approximation is justified with $0.3\%$ precision for $m_\nu=50\ {\rm meV}$ and $\langle p_0 \rangle=0.53\ {\rm meV}$. For $m_{\nu}>50\ {\rm meV}$, Eq.~(\ref{applambda}) becomes a more precise approximation than that for $m_\nu=50\ {\rm meV}$.
The present number density for non-relativistic neutrinos is given by
\begin{align}
n_{\nu}(t_0)\simeq e^{-t_0/\tau_{\nu}}f_cn_{\nu}^0.
\label{NDensity}
\end{align}

\begin{table}[h]
\begin{center}
	\begin{tabular}{cc}
		\hline \hline
		$m$ ({\rm meV}) & $f_c$   \\
		\hline 
		10 & 1.0053  \\
		50 & 1.12 \\
		100 & 1.5 \\
		200 & 3 \\
		\hline \hline
	\end{tabular}
	\caption{The enhancement factor, $f_c$, due to neutrino clustering by our Galaxy and nearby galaxies for given values of neutrino masses \cite{Mertsch:2019qjv}.}
  \label{tb:CND}
\end{center}
\end{table}

%%%%%%%%%%%%%%%%%%%%%%%%%%%%%%%%%%%%%%%%%%%%%%%%%%%%%%%%%%%%%%%%%%%%%%%%%%%%%%%%%%%%%%%%%%%%%%%%%%
%%%%%%%%%%%%%%%%%%%%%%%%%%%%%%%%%%%%%%%%%%%%%%%%%%%%%%%%%%%%%%%%%%%%%%%%%%%%%%%%%%%%%%%%%%%%%%%%%%
%%%%%%%%%%%%%%%%%%%%%%%%%%%%%%%%%%%%%%%%%%%%%%%%%%%%%%%%%%%%%%%%%%%%%%%%%%%%%%%%%%%%%%%%%%%%%%%%%%
%%%%%%%%%%%%%%%%%%%%%%%%%%%%%%%%%%%%%%%%%%%%%%%%%%%%%%%%%%%%%%%%%%%%%%%%%%%%%%%%%%%%%%%%%%%%%%%%%%

\section{Invisible neutrino decays}
\label{Sec3}

In this section, we discuss invisible neutrino decays induced by non-standard interactions of neutrinos. After giving the effective Lagrangians of non-standard neutrino interactions, we will estimate the decay rates of these channels.

In the following, we only consider 2-body and 3-body decays. Discussions on 4-body decays and beyond will be left to future work since such decay channels include many (unknown) parameters and we may not be able to impose severe constraints on these channels.

%%%%%%%%%%%%%%%%%%%%%%%%%%%%%%%%%%%%%%%%%%%%%%%%%%%%%%%%%%%%%%%%%%%%%%%%%%%%%%%%%%%%%%%%%%%%%%%%%%

\subsection{2-body decays}

In the case of 2-body decays, we consider the following Lagrangian at renormalizable level, which is the interaction between neutrinos $\nu_i$ and the pseudo-scalar bosons $\phi$,
\begin{align}
\mathcal{L}_{\rm int}=i\lambda_{ij}\phi \bar{\nu}_i\gamma^5\nu_j+{\rm h.c.}
\label{CBoson}
\end{align}
Here $\phi$ may be a majoron or an axion if an axion couples to neutrinos.
The indices $i, j\ (=1,2,3,4)$ denote mass eigenstates and $\lambda_{ij}$ are real coupling constants.
Note that $\nu_4$ denotes light sterile neutrinos (or unknown fermions) since heavier active neutrinos can decay into sterile neutrinos and bosons.
Although we can consider other interactions of 3-point coupling including scalar bosons or vector bosons, 
the constraints on these neutrino lifetimes will not be changed drastically, which is easily translated into the constraints on the couplings of other interactions.
Hereafter we only consider the interaction in Eq.~(\ref{CBoson}) as the case of 2-body decays.
We discuss other interactions of 2-body decays in appendix \ref{appa1}.
All of the decay rates in this work match with those in the latest version of ref. \cite{Escudero:2020ped}.
In the case of the interaction~(\ref{CBoson}), the rate of the 2-body decays, $\nu_i \rightarrow \nu_j \phi$, is

\begin{equation}
\Gamma_{\nu_i \rightarrow \nu_j  \phi}=\frac{\lambda_{ij}^2}{4\pi m_{\nu_i}}
\left[\left(1-\frac{m_{\nu_j}}{m_{\nu_i}}\right)^2-\frac{m_{\phi}^2}{m_{\nu_i}^2} \right]
\sqrt{\left[m_{\nu_i}^2-(m_{\nu_j}+m_{\phi})^2\right]\left[m_{\nu_i}^2-(m_{\nu_j}-m_{\phi})^2\right]} .
\end{equation}

\begin{figure}
	\begin{center}
	\includegraphics[clip,width=8.5cm]{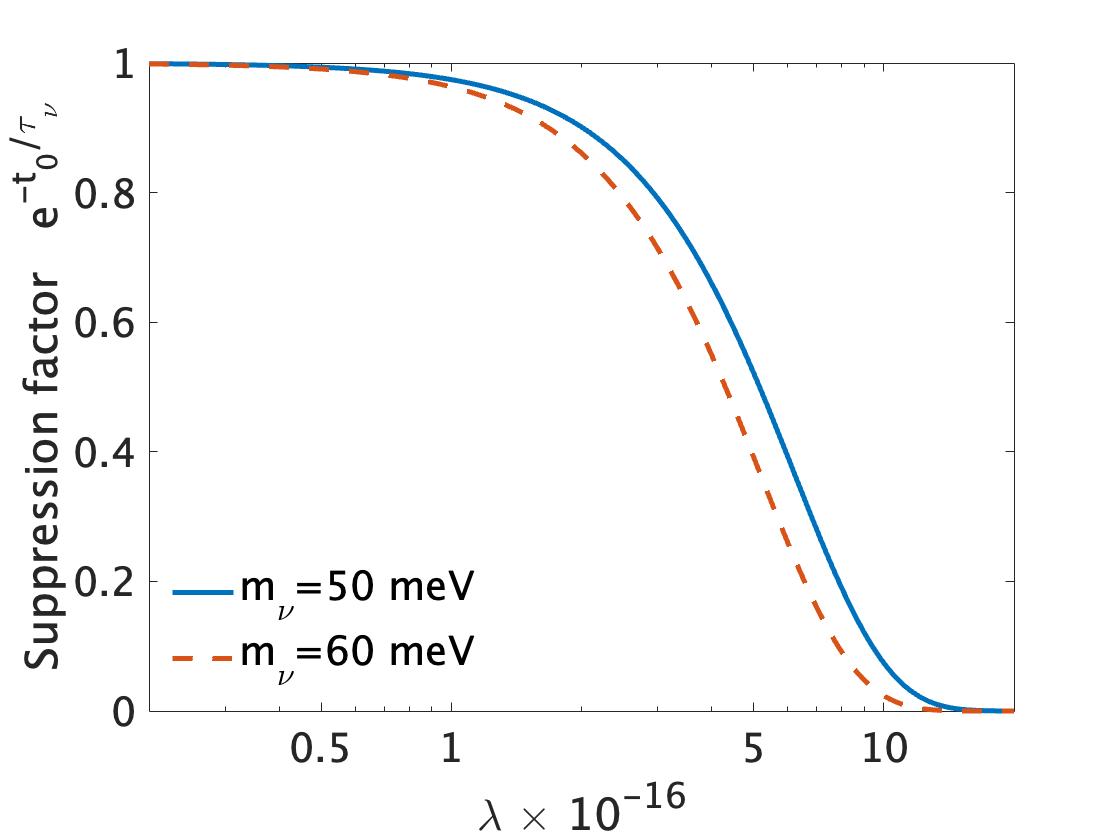}
	\end{center}
	 \vspace{-6mm}
	\caption{\small{Suppression factor, $e^{-t_0/\tau_{\nu_i}}$, as a function of $\lambda_{ij}$ for several values of $m_{\nu_i}$ in the case of Eq.~(\ref{ndecay2}).}}
	\label{fig:Sfactor_twobody}
\end{figure}

%\begin{align}
%\Gamma_{\nu_i \rightarrow \nu_j  %\phi}&=\frac{\lambda_{ij}^2}{4\pi %m_{\nu_i}}\sqrt{\left[m_{\nu_i}^2-(m_{\nu_j}+m_{\phi})^2\rig%ht]\left[m_{\nu_i}^2-(m_{\nu_j}-m_{\phi})^2\right]} %\nonumber \\
%&\ \ \ \ \  \times\left[\left(1-\frac{m_{\nu_j}}{m_{\nu_i}}\%right)^2-\frac{m_{\phi}^2}{m_{\nu_i}^2} \right].
%\end{align}
Here and hereafter we have assumed that neutrinos are Majorana fermions. For Dirac neutrinos, we need to replace $\lambda_{ij}$ with $\lambda_{ij}/2$.
In the case of $m_i \gg m_j, \ m_\phi$, we obtain
\begin{align}
\Gamma_{\nu_i \rightarrow \nu_j \phi}&\simeq \frac{\lambda_{ij}^2}{4\pi}m_{\nu_i} \nonumber \\
&= t_0^{-1}\left(\frac{m_{\nu_i}}{50\ {\rm meV}} \right)\left(\frac{\lambda_{ij}}{6.2\times 10^{-16}} \right)^2\,.
\end{align}
Then the present number density of $\nu_i$ is, according to $\tau_{\nu_i}=1/\Gamma_{\nu_i\rightarrow \nu_j \phi}$,
\begin{align}
n_{\nu_i}(t_0)
&\simeq \exp \left[-\left(\frac{m_{\nu_i}}{50\ {\rm meV}} \right)\left(\frac{\lambda_{ij}}{6.2\times 10^{-16}} \right)^2 \right]f_cn_{\nu_i}^0\,.
\label{ndecay2}
\end{align}
In Fig.~\ref{fig:Sfactor_twobody}, we plot the suppression factor, $e^{-t_0/\tau_{\nu_i}}$, as a function of $\lambda_{ij}$ for several values of $m_{\nu_i}$ in the case of Eq.~(\ref{ndecay2}).
%We may search the following parameter space of $\lambda_{ij}$, $0.2\times 10^{-15}<\lambda_{ij}< 2\times 10^{-15}$, in the future 100 gram tritium experiment.  
Thus, in the case of $m_i \gg m_j, \ m_\phi$, we can impose significant constraints on the coupling between neutrinos and bosons at the order of $10^{-15}$ via future direct detections of the C$\nu$B.

%%%%%%%%%%%%%%%%%%%%%%%%%%%%%%%%%%%%%%%%%%%%%%%%%%%%%%%%%%%%%%%%%%%%%%%%%%%%%%%%%%%%%%%%%%%%%%%%%%

\subsection{3-body decays}
If pseudo-scalar bosons are heavier than neutrinos, $m_{\phi}>m_{\nu_i}$, the 2-body decays in the previous section are kinematically forbidden.
Even in this case, 3-body decays of neutrinos are induced by mediating pseudo-scalar bosons as off-shell particles,
$\nu_i \rightarrow \nu_j\phi \rightarrow \nu_j \nu_k \bar{\nu_l}$ if $m_{\nu_i}>m_{\nu_j}+m_{\nu_k}+m_{\nu_l}$.
The effective Lagrangian that causes the 3-body decays can be written from Eq.~(\ref{CBoson}) as
\begin{align}
\mathcal{L}_{\rm eff}=G_{\rm eff}\left[\bar{\nu}_i\gamma^5\nu_j\bar{\nu}_k\gamma^5\nu_l+\bar{\nu}_i\gamma^5\nu_j\bar{\nu}_l\gamma^5\nu_k+ {\rm h.c.}  \right],
\label{Fermieff}
\end{align}
where
\begin{align}
G_{\rm eff}=\frac{\lambda_{ij}\lambda_{kl}}{m_{\phi}^2}.
\end{align}
For Majorana neutrinos, the decay rate of $\nu_i\rightarrow \nu_j \nu_k \bar{\nu}_l$ is
\begin{align}
\Gamma_{\nu_i\rightarrow \nu_j \nu_k \bar{\nu}_l}
&\simeq \frac{G_{\rm eff}^2}{3\cdot2^5\pi^3}m_{\nu_i}^5, \nonumber \\
&= t_0^{-1}\left(\frac{m_{\nu_i}}{50\ {\rm meV}} \right)^5\left(\frac{G_{\rm eff}}{3.8\times 10^{-12}\ {\rm eV}^{-2}} \right)^2,
\label{Threedecay}
\end{align}
where we assume $m_{\nu_i} \gg m_{\nu_j}, m_{\nu_k}, m_{\nu_l}$ and $j\neq k$. For $j=k$, we need an additional factor of $1/2$.
Note that we cannot find the exact analytic formula of this decay rate at tree level although we can get the analytic formulae for some specific cases. See also appendix \ref{appa2}, where the decay rates match with those in the latest version of  ref.~\cite{Escudero:2020ped}.

\begin{figure}
	\begin{center}
	\includegraphics[clip,width=8.5cm]{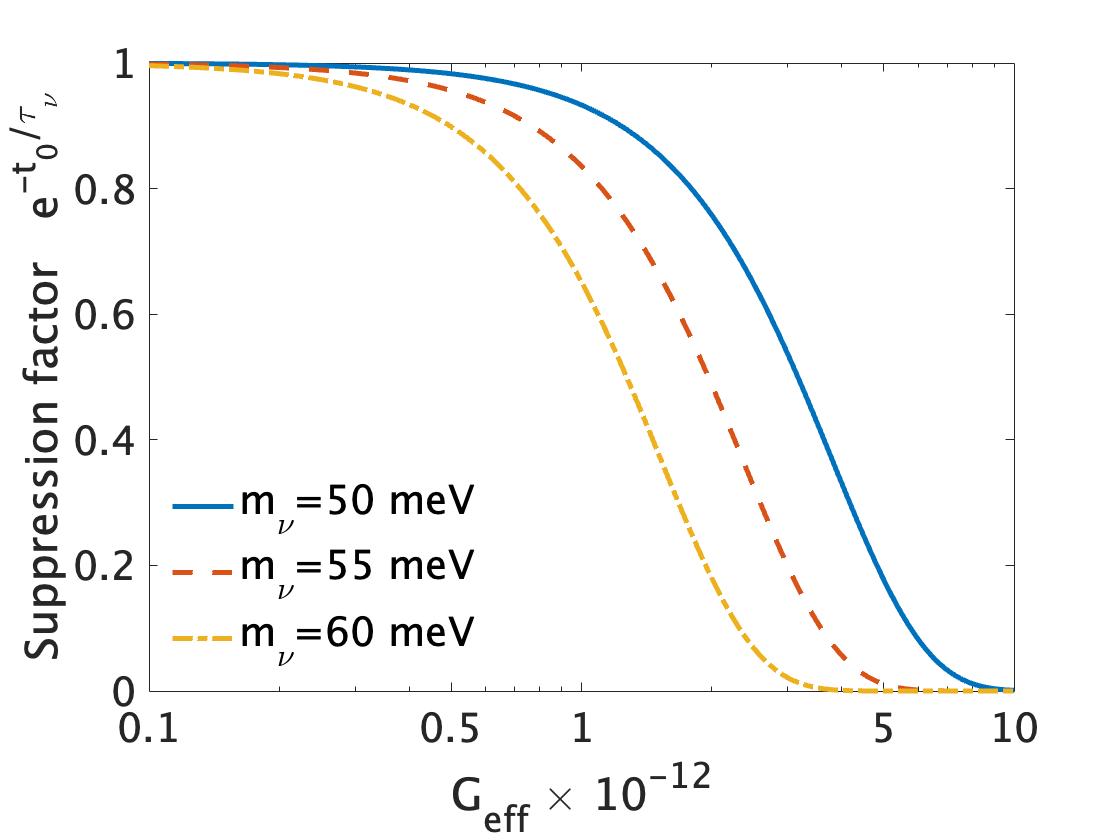}
	\end{center}
	 \vspace{-6mm}
	\caption{\small{Suppression factor, $e^{-t_0/\tau_{\nu_i}}$, as a function of $\lambda_{ij}$ for several values of $m_{\nu_i}$ in the case of Eq.~(\ref{Threedecay}).}}
	\label{fig:Sfactor_threebody}
\end{figure}

In Fig.~\ref{fig:Sfactor_threebody}, we plot the suppression factor, $e^{-t_0/\tau_{\nu_i}}$, as a function of $G_{\rm eff}$ for several values of $m_{\nu_i}$
in the case of Eq.~(\ref{Threedecay}). In the case of $m_{\nu_i} \gg m_{\nu_j}, m_{\nu_k}, m_{\nu_l}$, we can impose significant constraints on the effective coupling, $G_{\rm eff}$, at the order $10^{-12}\ {\rm eV}^{-2}$ in future direct detections of cosmic neutrinos. 

%%%%%%%%%%%%%%%%%%%%%%%%%%%%%%%%%%%%%%%%%%%%%%%%%%%%%%%%%%%%%%%%%%%%%%%%%%%%%%%%%%%%%%%%%%%%%%%%%%
%%%%%%%%%%%%%%%%%%%%%%%%%%%%%%%%%%%%%%%%%%%%%%%%%%%%%%%%%%%%%%%%%%%%%%%%%%%%%%%%%%%%%%%%%%%%%%%%%%
%%%%%%%%%%%%%%%%%%%%%%%%%%%%%%%%%%%%%%%%%%%%%%%%%%%%%%%%%%%%%%%%%%%%%%%%%%%%%%%%%%%%%%%%%%%%%%%%%%
%%%%%%%%%%%%%%%%%%%%%%%%%%%%%%%%%%%%%%%%%%%%%%%%%%%%%%%%%%%%%%%%%%%%%%%%%%%%%%%%%%%%%%%%%%%%%%%%%%

\section{Lighter neutrino spectra from heavier neutrino decays}
\label{Sec4}

In this section, we discuss the expected spectra of lighter neutrinos produced by the decays of heavier neutrinos.
We begin to review generic arguments of the expected spectra of particles produced by the decays of heavy particles. Then, we give concrete formulae of the spectra in the 2-body and 3-body invisible decays for neutrinos.

%%%%%%%%%%%%%%%%%%%%%%%%%%%%%%%%%%%%%%%%%%%%%%%%%%%%%%%%%%%%%%%%%%%%%%%%%%%%%%%%%%%%%%%%%%%%%%%%%%
%%%%%%%%%%%%%%%%%%%%%%%%%%%%%%%%%%%%%%%%%%%%%%%%%%%%%%%%%%%%%%%%%%%%%%%%%%%%%%%%%%%%%%%%%%%%%%%%%%
\subsection{Spectra from neutrino decays}

In this section, we derive the expected spectrum of lighter neutrinos produced by the decays of heavier neutrinos.

The decay of one particle of heavier neutrino $\nu_i$ injects $N_{\nu_j}$ numbers of lighter neutrinos $\nu_j$ with a spectrum $dN/dE_{\nu_j}$,
\begin{align}
\frac{dN}{dE_{\nu_j}}=\frac{N_{\nu_j}}{\Gamma_{\nu}}\frac{d\Gamma_{\nu} }{dE_{\nu_j}},
\ \ \ \ \int dE_{\nu_j}\frac{dN}{dE_{\nu_j}}=N_{\nu_j},
\label{Onespectrum}
\end{align}
where $N(t)$ denotes the particle number of lighter neutrinos $\nu_j$ from the decay of one particle of heavier neutrino $\nu_i$.
$N(t)=0$ before the decay of $\nu_i$ while $N(t)=N_{\nu_j}$ after the decay of $\nu_i$. 

In the following, we take into account the energy dilution due to the cosmic expansion and the superposition from non-instantaneous decays of neutrinos. The number density
of the lighter neutrinos $\nu_j$ produced by the decays of the heavier neutrinos $\nu_i$ 
 is written from Eq.~(\ref{NDensity}) as  
\begin{align}
\tilde{n}_{\nu_j}(t)&=N_{\nu_j}n_{\nu_i}^0\left[\frac{a(t_0)}{a(t)} \right]^3\left(1-e^{-\int_{t_d}^{t} \frac{dt'}{\tau_{\nu}'(t')}}\right) \nonumber \\
&\simeq N_{\nu_j}n_{\nu_i}^0\frac{1-e^{-t/\tau_{\nu_i}}}{a(t)^3},
\label{NDproduced}
\end{align}
where $a(t)$ is the scale factor of the universe with $a(t_0)=1$ and we assume that $t$ is sufficiently large so that heavier neutrinos are non-relativistic at the time $t$.
From Eq.~(\ref{NDproduced}), the comoving number density of lighter neutrinos can be written as
\begin{align}
\frac{d\left(\tilde{n}_{\nu_j}a^3\right)}{dt}=\frac{N_{\nu_j}n_{\nu_i}^0}{\tau_{\nu_i}}e^{-t/\tau_{\nu_i}}.
\end{align}
After changing the time $t$ to the redshift $z$ and doing a little calculation, we obtain
\begin{align}
\frac{d\left(\tilde{n}_{\nu_j}a^3 \right)}{dE_{\nu_j}}dN=-\frac{1}{1+z}\frac{N_{\nu_j}n_{\nu_i}^0}{\tau_{\nu_i}}e^{-t(z)/\tau_{\nu_i}}\frac{1}{H(z)}\frac{dN}{dE_{\nu_j}}dz,
\end{align}
where $\int dz \frac{dN}{dz}=-N_{\nu_j}$.
In addition, $dN/dE_{\nu_j}$ can be rewritten as
\begin{align}
\frac{dN}{dE_{\nu_j}}&=(1+z)\frac{E_{\nu_j}}{p_{\nu_j}}\frac{p_{\nu_j}(z)}{E_{\nu_j}(z)}\frac{dN}{dE_{\nu_j}(z)},
\end{align}
where 
\begin{align}
p_{\nu_j}(z)&=p_{\nu_j}(1+z), \nonumber \\
E_{\nu_j}(z)&=\sqrt{p_{\nu_j}^2(1+z)^2+m_{\nu_j}^2}.
\end{align}
$p_{\nu_j}(z)$ and $E_{\nu_j}(z)$ denote the momentum and the energy at the redshift $z$ to which $p_{\nu_j}$ and $E_{\nu_j}$ in the current detection are extrapolated (blue-shifted).
Finally, after integrating from $z=0$ to $z=z_d\simeq \infty$, we obtain the present energy spectrum of the lighter neutrinos $\nu_j$,
\begin{align}
\frac{d\tilde{n}_{\nu_j}^0}{dE_{\nu_j}}=\frac{n_{\nu_i}^0}{\tau_{\nu_i}}\int^{\infty}_0dz~e^{-t(z)/\tau_{\nu_i}}\frac{1}{H(z)}\frac{E_{\nu_j}}{p_{\nu_j}}\frac{p_{\nu_j}(z)}{E_{\nu_j}(z)}\frac{dN}{dE_{\nu_j}(z)}(E_{\nu_j}(z)),
\label{ST}
\end{align}
where $\tilde{n}_{\nu_j}^0 \equiv \tilde{n}_{\nu_j}(t_0) = \tilde{n}_{\nu_j}(t_0)a(t_0)^3$.
We have also derived Eq.~(\ref{ST}) from the Boltzmann equation for neutrinos (see e.g. ref.~\cite{Fogli:2004gy}).
In the epoch when the universe is mainly composed of dark matter and dark energy, the Hubble parameter is approximately given by
\begin{align}
H(z)\simeq H_0\sqrt{\Omega_m(1+z)^3+\Omega_\Lambda}.
\end{align}
This approximation is valid when we search for the neutrino decays in the universe dominated by dark matter and dark energy.
Under this approximation, we can express the time of the universe at redshift $z$ as
\begin{align}
t(z)\simeq\frac{2}{3H_0\sqrt{\Omega_\Lambda}}\ln \left(\sqrt{r(z)}+\sqrt{1+r(z)} \right),
\label{tz}
\end{align}
where $r(z)$ is given by
\begin{align}
r(z)=\frac{\Omega_\Lambda}{\Omega_m}\frac{1}{(1+z)^3}.
\end{align}

Finally we comment on effect of gravitational attraction by our Galaxy on Eq.~(\ref{ST}).
From Eq.~(\ref{ST}), the superposition of neutrino decays around $z\sim 1$ is mainly contributing to the present spectrum of the lighter neutrinos since Eq.~(\ref{ST}) is suppressed by $H(z)$ at $z\gg 1$ and $dN/dE(z)$ is normalized. At $z\sim 1$, neutrinos decay outside our Galaxy. Then the heavier neutrinos are not clustered. In addition, the lighter neutrinos produced by decays are more energetic than neutrinos produced in the early universe due to energy injections by decays, resulting in higher velocity of neutrinos produced by decays.
Thus, we expect that the lighter neutrino spectra with mass $m$ produced by the decays of the heavier neutrinos are less clustered by our Galaxy, compared with neutrinos produced in the early universe as in Table~\ref{tb:CND}.
We leave this precise estimation to future work.
%%%%%%%%%%%%%%%%%%%%%%%%%%%%%%%%%%%%%%%%%%%%%%%%%%%%%%%%%%%%%%%%%%%%%%%%%%%%%%%%%%%%%%%%%%%%%%%%%%
%%%%%%%%%%%%%%%%%%%%%%%%%%%%%%%%%%%%%%%%%%%%%%%%%%%%%%%%%%%%%%%%%%%%%%%%%%%%%%%%%%%%%%%%%%%%%%%%%%

\subsection{2-body decays}
In a 2-body decay $\nu_i\rightarrow \nu_j \phi$, a lighter neutrino $\nu_j$ has the following monochromatic momentum and energy at the decay,
\begin{align}
p^{\ast}&=\frac{1}{2m_{\nu_i}}\sqrt{[m_{\nu_i}^2-(m_{\nu_j}+m_\phi)^2][m_{\nu_i}^2-(m_{\nu_j}-m_\phi)^2]}, \nonumber \\
E^{\ast}&=\frac{m_{\nu_i}^2+m_{\nu_j}^2-m_{\phi}^2}{2m_{\nu_i}}.
\label{East}
\end{align}
Then Eq.~(\ref{Onespectrum}) in a 2-body decay becomes a monochromatic spectrum,
\begin{align}
\frac{dN}{dE_{\nu_j}}\biggl|_{2-{\rm body}}=\delta \left(E_{\nu_j}- E^{\ast}\right),
\label{Sonetwo}
\end{align}
where $N_{\nu_j}=1$.
We can obtain the expected spectrum of lighter neutrinos in the case of 2-body decays after substituting Eq.~(\ref{Sonetwo}) into Eq.~(\ref{ST}).
However, in the following, we will rederive it in another simple way, following ref.~\cite{McKeen:2018xyz}.

At the time of the production of a lighter neutrino, $t_{E}\equiv t(z_E)$, it has a momentum of $p^\ast$. Owing to the expansion of the universe, the momentum of the lighter neutrino is $p_{\nu_j}=p^\ast a(t_E)$. Thanks to this one-to-one correspondence between the current momentum and that at the production, 
the present energy spectrum of the lighter neutrinos is related to the comoving number density of neutrinos,
\begin{align}
\frac{d\tilde{n}_{\nu_j}^0}{dE_{\nu_j}}
&=\frac{d(a(t)^3\tilde{n}_{\nu_j}(t))}{dE_{\nu_j}} \biggl|_{t=t_E} \nonumber \\
&=\frac{1}{p^\ast}\frac{E_{\nu_j}}{p_{\nu_j}}\frac{d(a(t)^3\tilde{n}_{\nu_j}(t))}{da(t)} \biggl|_{t=t_E} \nonumber \\
&=n_{\nu_i}^0\frac{e^{-t(z_E)/\tau_{\nu_i}}}{H(z_E)\tau_{\nu_i}}\frac{E_{\nu_j}}{E_{\nu_j}^2-m_{\nu_j}^2}.
\label{Stwo}
\end{align}
where $t_E = t(z_E)$ and $H(z_E)$ are the time and the Hubble parameter at the decay, respectively. 
For the lighter neutrinos with present energy $E_{\nu_j}$ produced by the decay of the heavy neutrino, the redshift at the decay $z_E$ is given by
\begin{align}
1+z_E&=\frac{p^\ast}{\sqrt{E_{\nu_j}^2-m_{\nu_j}^2}}.
\end{align}

%%%%%%%%%%%%%%%%%%%%%%%%%%%%%%%%%%%%%%%%%%%%%%%%%%%%%%%%%%%%%%%%%%%%%%%%%%%%%%%%%%%%%%%%%%%%%%%%%%
%%%%%%%%%%%%%%%%%%%%%%%%%%%%%%%%%%%%%%%%%%%%%%%%%%%%%%%%%%%%%%%%%%%%%%%%%%%%%%%%%%%%%%%%%%%%%%%%%%

\subsection{3-body decays}
In order to obtain the energy spectrum in the case of 3-body decays, $\nu_i\rightarrow \nu_j\nu_k\bar{\nu_l}$, we consider $d\Gamma_{\nu_i\rightarrow \nu_j\nu_k\bar{\nu_l}}/ dE_{\nu_j}$ and $\Gamma_{\nu_i\rightarrow \nu_j\nu_k\bar{\nu_l}}$ in Eq.~(\ref{Onespectrum}).
The decay rate $\Gamma_{\nu_i\rightarrow \nu_j\nu_k\bar{\nu_l}}$ is given by
\begin{align}
\Gamma_{\nu_i\rightarrow \nu_j\nu_k\bar{\nu_l}}&=\frac{1}{2^9\pi^5m_{\nu_i}}\int\frac{d^3p_{\nu_j} d^3p_{\nu_k} d^3p_{\nu_l}}{E_{\nu_j}E_{\nu_k}E_{\nu_l}}|\mathcal{M}|^2 \delta^4(p_{\nu_i}-p_{\nu_j}-p_{\nu_k}-p_{\nu_l}) \nonumber \\
&=\frac{1}{2^6\pi^3m_{\nu_i}}\int dE_{\nu_j} dE_{\nu_k}|\mathcal{M}|^2,
\label{Gamma3}
\end{align}
where $|\mathcal{M}|^2$ is the squared matrix element averaged over spins for the initial state and summed over spins for the final states. Here we assume $j\neq k$, and for $j=k$, we need an additional factor of $1/2$.
Next we integrate over $E_{\nu_k}$ for each $E_{\nu_j}$ in Eq.~(\ref{Gamma3}) and $d\Gamma_{\nu_i\rightarrow \nu_j\nu_k\bar{\nu_l}}/ dE_{\nu_j}$.
The lower (upper) limit of the integration denotes $E_{\nu_k}^{\rm min}\ (E_{\nu_k}^{\rm max})$.
After some calculations, $E_{\nu_k}^{\rm max}-E_{\nu_k}^{\rm min}$ is given by
\begin{align}
E_{\nu_k}^{\rm max}-E_{\nu_k}^{\rm min}&=\frac{2m_{\nu_i} |\bm{p}_{\nu_j}|}{m_{\nu_i}^2-2m_{\nu_i}E_{\nu_j}+m_{\nu_j}^2}(E_{\nu_j}^{\rm max}-E_{\nu_j})^{1/2}\left(E_{\nu_j}^{\rm max}-E_{\nu_j}+\frac{2m_{\nu_k}m_{\nu_l}}{m_{\nu_i}}\right)^{1/2}.
\label{minmax}
\end{align}
$E_{\nu_j}^{\rm max}$ is the maximal value of $E_{\nu_j}$ in the 3-body decay, which is given by 
\begin{align}
E_{\nu_j}^{\rm max}=\frac{1}{2m_{\nu_i}}\left[m_{\nu_i}^2+m_{\nu_j}^2-\left(m_{\nu_k}+m_{\nu_l}\right)^2\right].
\label{Emax3}
\end{align}
Then $d\Gamma_{\nu_i\rightarrow \nu_j\nu_k\bar{\nu_l}}/ dE_{\nu_j}$ is given by
\begin{align}
\frac{d\Gamma_{\nu_i\rightarrow \nu_j\nu_k\bar{\nu_l}}}{dE_{\nu_j}}=\frac{1}{2^6\pi^3m_{\nu_i}}\int^{E_{\nu_k}^{\rm max}}_{E_{\nu_k}^{\rm min}} dE_{\nu_k}|\mathcal{M}|^2.
\end{align}
For the case of Majorana neutrinos, we obtain $|\mathcal{M}|^2$ in the 3-body decays mediated by pseudo-scalar bosons,
\begin{align}
|\mathcal{M}|^2=2^7G_{\rm eff}^2(m_{\nu_i}E_{\nu_j}-m_{\nu_i}m_{\nu_j})(E_{\nu_k}E_{\nu_l}-\bm{p}_{\nu_k}\cdot\bm{p}_{\nu_l}+m_{\nu_k}m_{\nu_l}),
\end{align}
where
\begin{align}
\bm{p}_{\nu_k}\cdot\bm{p}_{\nu_l}&=\frac{1}{2}\left[2E_{\nu_k}E_{\nu_l}+2m_{\nu_i}E_{\nu_j}+m_{\nu_k}^2+m_{\nu_l}^2-m_{\nu_i}^2-m_{\nu_j}^2\right].
\end{align}
Finally, we obtain
 \begin{align}
 \frac{d\Gamma_{\nu_i\rightarrow \nu_j\nu_k\bar{\nu_l}}}{dE_{\nu_j}}&=\frac{2G_{\rm eff}^2}{\pi^3}\frac{m_{\nu_i}|\bm{p}_{\nu_j}|(E_{\nu_j}-m_{\nu_j})}{m_{\nu_i}^2-2m_{\nu_i}E_{\nu_j}+m_{\nu_j}^2}
 \left[m_{\nu_i}^2+m_{\nu_j}^2-(m_{\nu_k}-m_{\nu_l})^2-2m_{\nu_i}E_{\nu_j} \right] \nonumber \\
 & \ \ \ \ \  \times \sqrt{(E_{\nu_j}^{\rm max}-E_{\nu_j})\left(E_{\nu_j}^{\rm max}-E_{\nu_j}+\frac{2m_{\nu_k}m_{\nu_l}}{m_{\nu_i}}\right)}.
 \end{align}
 $d\Gamma_{\nu_i\rightarrow \nu_j\nu_k\bar{\nu_l}}/ dE_{\nu_j}$ obviously vanishes at $E_{\nu_j}=E_{\nu_j}^{\rm max}$ since the value of $E_{\nu_{j}}$ is kinematically allowed from $m_{\nu_j}$ to $E_{\nu_j}^{\rm max}$.
 We also obtain the decay rate $\Gamma_{\nu_i \rightarrow \nu_j \nu_k \bar{\nu}_l}$ as 
 \begin{align}
 \Gamma_{\nu_i \rightarrow \nu_j\nu_k\bar{\nu}_l}=\int^{E_{\nu_j}^{\rm max}}_{m_{\nu_j}} dE_{\nu_j} \frac{d\Gamma_{\nu_i\rightarrow \nu_j\nu_k\bar{\nu}_l}}{dE_{\nu_j}}.
 \end{align}
 Note that the formula of $|\mathcal{M}|^2$ depends on the mediated bosons. In appendix~\ref{appa2}, we show $d\Gamma_{\nu_i\rightarrow \nu_j\nu_k\bar{\nu_l}}/ dE_{\nu_j}$ in the cases of mediated scalar and vector bosons.
 Note also that $dN/dE_{\nu_j}$ in Eq.~(\ref{Onespectrum}) is independent of the coupling constant, $G_{\rm eff}$.

%%%%%%%%%%%%%%%%%%%%%%%%%%%%%%%%%%%%%%%%%%%%%%%%%%%%%%%%%%%%%%%%%%%%%%%%%%%%%%%%%%%%%%%%%%%%%%%%%%
%%%%%%%%%%%%%%%%%%%%%%%%%%%%%%%%%%%%%%%%%%%%%%%%%%%%%%%%%%%%%%%%%%%%%%%%%%%%%%%%%%%%%%%%%%%%%%%%%%
%%%%%%%%%%%%%%%%%%%%%%%%%%%%%%%%%%%%%%%%%%%%%%%%%%%%%%%%%%%%%%%%%%%%%%%%%%%%%%%%%%%%%%%%%%%%%%%%%%
%%%%%%%%%%%%%%%%%%%%%%%%%%%%%%%%%%%%%%%%%%%%%%%%%%%%%%%%%%%%%%%%%%%%%%%%%%%%%%%%%%%%%%%%%%%%%%%%%%

\section{Forecasts of constraints in neutrino capture experiments on tritium}
\label{Sec5}

Cosmic neutrinos can be captured on tritium with a half-life of $12.32$ years by the following process,
\begin{align}
\nu_i+\mathrm{^3H}\rightarrow \mathrm{^3He}+e^-.
\label{Inversebeta}
\end{align}
In this section, we estimate the would-be observed spectrum of the emitted electrons and their detectability from the neutrino decays in future experiments of neutrino capture on tritium.
In particular, we focus on the PTOLEMY-type experiment, assuming $100$ g of tritium as a target.

In the case of Majorana neutrinos, the capture rate of cosmic neutrinos $\nu_i$ with the number density $n_{\nu_i}$ on tritium is simply given by  \cite{Long:2014zva} \footnote{In the case of non-relativistic Dirac neutrinos, the capture rate is given by $\Gamma_{\rm C\nu B}^i\simeq|U_{ei}|^2 \bar{\sigma}n_{\nu_i} N_T$ \cite{Long:2014zva}. For relativistic Dirac neutrinos, the capture rate is discussed in refs.~\cite{Roulet:2018fyh,Akita:2020jbo}}
\begin{align}
\Gamma_{\rm C\nu B}^i&\simeq2|U_{ei}|^2 \bar{\sigma} n_{\nu_i} N_T, \nonumber \\
\bar{\sigma}&=3.76\times 10^{-45}\ {\rm cm^2},
\end{align}
where $N_T=\mathrm{M_T}/m_{\mathrm{^3H}}$ is the number of tritium, $\mathrm{M_T}$ is the total mass of the experimental setup of tritium, and $M_{\mathrm{^3H}}\simeq 2809.432\ {\rm MeV}$ is the atomic mass of the $\mathrm{^3H}$ \cite{Wang2016}.
$|U_{ei}|^2\bar{\sigma}$ is the cross section of Eq.~(\ref{Inversebeta}), which is approximately constant, and $|U_{ei}|$ is an element of the Pontecorvo-Maki-Nakagawa-Sakata (PMNS) matrix, which is measured to be \cite{Esteban:2020cvm}
\begin{align}
|U_{e1}|^2\simeq 0.681 ,\ \ \ \ |U_{e2}|^2\simeq 0.297,\ \ \ \ |U_{e3}|^2\simeq 0.0222.
\label{PMNS}
\end{align}
The capture rate on tritium of lighter neutrinos $\nu_j$ produced by the neutrino decays and the spectrum of the electrons emitted in this process are given by
\begin{align}
\tilde{\Gamma}_{\rm C\nu B}^j&\simeq2|U_{ej}|^2 \bar{\sigma} \tilde{n}_{\nu_j}^0 N_T \nonumber \\
&\simeq 2|U_{ej}|^2 \bar{\sigma} n_{\nu_i}^0 N_T~(1-e^{-t_0/\tau_{\nu_i}}), \nonumber \\
\frac{d\tilde{\Gamma}_{\rm C\nu B}^j}{dE_e}&=2|U_{ej}|^2\bar{\sigma} N_T\frac{d\tilde{n}_{\nu_j}^0}{dE_e}.
\end{align}
From the energy-momentum conservation, the energy of the emitted electron, $E_e$, is \cite{Long:2014zva}
\begin{align}
E_e&\simeq K_{\rm end}^0+m_e+E_{\nu_j}, \nonumber \\
K_{\rm end}^0&=\frac{(m_{\mathrm{^3H}}-m_e)^2-m_{\mathrm{^3He}}^2}{m_{\mathrm{^3H}}}.
\label{relationEeEnu}
\end{align}
We can rewrite the emitted electron spectrum from the lighter neutrinos $\nu_j$ produced by the neutrino decays as
\begin{align}
\frac{d\tilde{\Gamma}_{\rm C\nu B}^j}{dE_e}(E_e)=2|U_{ej}|^2\bar{\sigma} N_T\frac{d\tilde{n}_{\nu_j}^0}{dE_{\nu_j}}(E_e).
\label{Spectrumdecay}
\end{align}

%%%%%%%%%%%%%%%%%%%%%%%%%%%%%%%%%%%%%%%%%%%%%%%%%%%%%%%%%%%%%%%%%%%%%%%%%%%%%%%%%%%%%%%%%%%%%%%%%%
\subsection{Observed spectra for unstable neutrinos}
\label{Sec5.1}

In this section, we show the would-be observed spectrum of an electron emitted from the process given in Eq.~(\ref{Inversebeta}) in the case that cosmic neutrinos decay into lighter neutrinos until today via the interactions of Eqs.~(\ref{CBoson}) or (\ref{Fermieff}).
One of the main challenges for detecting the signals of C$\nu$B is to distinguish them from the background of tritium $\beta$-decay, $\mathrm{^3H}\rightarrow \mathrm{^3He}+e^-+\bar{\nu}_i$. 
The $\beta$-decay spectrum near the endpoint is given by \cite{Masood:2007rc}
\begin{align}
\frac{d\Gamma_\beta}{dE_e}=\frac{\bar{\sigma}}{\pi^2}N_T\sum_{i=1}^{3}|U_{ei}|^2H(E_e,m_{\nu_i}),
\end{align}
where
\begin{align}
H(E_e, m_{\nu_i}) = \frac{1-m_e^2/(E_em_{\mathrm{^3H}})}{(1-2E_e/m_{\mathrm{^3H}}+m_e^2/m_{\mathrm{^3H}}^2)^2}
\sqrt{y_i\left(y_i+\frac{2m_{\nu_i}m_{\mathrm{^3He}}}{m_{\mathrm{^3H}}}\right)} \left[y_i+\frac{m_{\nu_i}}{m_{\mathrm{^3H}}}(m_{\mathrm{^3He}}+m_{\nu_i})\right],
\end{align}
with $y_i\simeq E_{\rm end}^i-E_e$. Here $E_{\rm end}^i$ denotes the maximum electron energy in the process of $\mathrm{^3H}\rightarrow \mathrm{^3He}+e^-+\bar{\nu_i}$, 
\begin{align}
E_{\rm end}^i\simeq K_{\rm end}^0+m_e-m_{\nu_i}.
\end{align}
$E_{\rm end}^i$ for the lightest neutrinos is the maximum electron energy in the $\beta$-decay called the $\beta$-decay endpoint energy,
\begin{align}
    E_{\rm end}\simeq K_{\rm end}^0+m_e-m_{\rm lightest}.
    \label{endpoint}
\end{align}

To distinguish the signals of C$\nu$B from the $\beta$-decay background, we need to take the energy resolution of the detector $\Delta$ into account.
We model the Gaussian-smeared spectrum as an actual spectrum.
Then the Gaussian-smeared $\beta$-decay spectrum is given by
\begin{align}
\frac{d\bm{\Gamma}_\beta}{dE_e}=\frac{1}{\sqrt{2\pi}\sigma}\int^{\infty}_{-\infty}dE_e'\frac{d\Gamma_{\beta}}{dE_e}(E_e')\exp\left[-\frac{(E_e'-E_e)^2}{2\sigma^2} \right],
\end{align}
where $\sigma=\Delta/\sqrt{8\ln2}$ is a standard deviation, not a cross section.
The Gaussian-smeared spectrum of the electrons from cosmic neutrinos produced in the early universe, $d\bm{\Gamma}_{\nu_i}/dE_e$, and that of the electrons from lighter neutrinos produced by the decays of heavier neutrinos, $d\tilde{\bm{\Gamma}}_{\nu_j}/dE_e$ , are given respectively by\footnote{In the following sections \ref{Sec5.1} and \ref{Sec5.2}, when neutrino masses are degenerate ($m_{\rm lightest}\geq 50\ {\rm meV}$), we assume for simplicity that the number densities for all neutrino species produced in the early universe are enhanced by the gravitational clustering as much as the same amount as that of the lightest neutrinos.}
\begin{align}
\frac{d\bm{\Gamma}_{\rm C\nu B}^i}{dE_e}&=\frac{1}{\sqrt{2\pi}\sigma}\int^{\infty}_{-\infty} dE_e'\ \Gamma_{\rm C\nu B}^i(E_e') \delta[E_e'-(E_{\rm end}+E_{\nu_i}+m_{\rm lightest})]\exp \left[-\frac{(E_e'-E_e)^2}{2\sigma^2} \right], \nonumber \\
\frac{d\tilde{\bm{\Gamma}}_{\rm C\nu B}^j}{dE_e}&=\frac{1}{\sqrt{2\pi}\sigma}\int^{\infty}_{-\infty}dE_e'\ \frac{d\tilde{\Gamma}_{\rm C\nu B}^j}{dE_e}(E_e') \exp \left[-\frac{(E_e'-E_e)^2}{2\sigma^2} \right].
\end{align}

The actual expected spectrum is a superposition of $d\bm{\Gamma}_\beta/dE_e,\ d\bm{\Gamma}_{\rm C\nu B}^i/dE_e$ and $d\tilde{\bm{\Gamma}}_{\rm C\nu B}^i/dE_e$.
However, showing each spectrum separately would give us a better prospect for understanding the nature of neutrino decays.
Then, in the following, we will show the expected electron spectra, $d\bm{\Gamma}_\beta/dE_e,\ d\bm{\Gamma}_{\rm C\nu B}^i/dE_e$ and $d\tilde{\bm{\Gamma}}_{\rm C\nu B}^i/dE_e$, separately.

%%%%%%%%%%%%%%%%%%%%%%%%%%%%%%%%%%%%%%%%%%%%%%%%%%%%%%%%%%%%%%%%%%%%%%%%%%%%%%%%%%%%%%%%%%%%%%%%%%
\subsubsection{2-body decays}

In the following, we consider the expected spectra in the case of 2-body decays, assuming $100$ g of a tritium target.
Due to the ambiguities of the couplings to invisible particles and the neutrino masses, we will not explore all of the possible parameter regions. Hereafter we consider only the decay processes of $\nu_3 \rightarrow \nu_1 \phi$ in the normal mass ordering and $\nu_1 \rightarrow \nu_3 \phi$ in the inverted mass ordering. This assumption implies that we set $\lambda_{ij}$ in Eq.~(\ref{CBoson}) as
\begin{align}
{\rm NO}&:\ \ \lambda_{13} \neq 0\ \ {\rm and}\ \ \lambda_{ij}=0\ (i\neq 1,\ j\neq 3), \nonumber \\
{\rm IO}&:\ \  \lambda_{31} \neq 0\ \ {\rm and}\ \ \lambda_{ij}=0\ (i\neq 3,\ j\neq 1). \nonumber 
\end{align}
In this setup, $\nu_2$ does not decay and is not produced by the decays of heavier neutrinos. In addition, we take the neutrino masses to satisfy the observed values of neutrino squared-mass differences from neutrino oscillation experiments \cite{Esteban:2020cvm}, 
$\Delta m_{21}^2 \simeq (8.6\ {\rm meV})^2$ and $ |\Delta m_{3l}^2|  \simeq (50\ {\rm meV})^2$.
We take $l=1$ for the NO case and $l=2$ for the IO case as in ref. \cite{Esteban:2020cvm}.

\begin{figure}[htbp]
 \begin{minipage}{0.5\hsize}
  \begin{center}
   \includegraphics[clip,width=85mm]{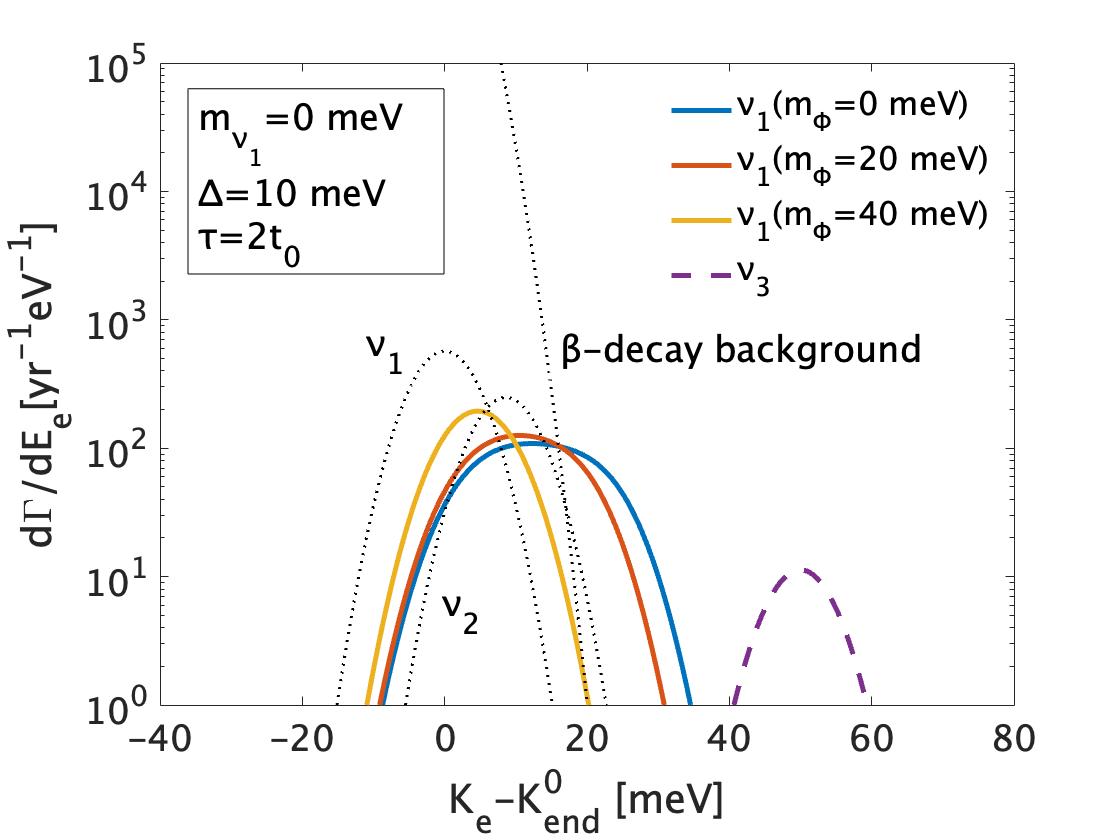}
  \end{center}
 \end{minipage}
 \begin{minipage}{0.5\hsize}
  \begin{center}
   \includegraphics[clip,width=85mm]{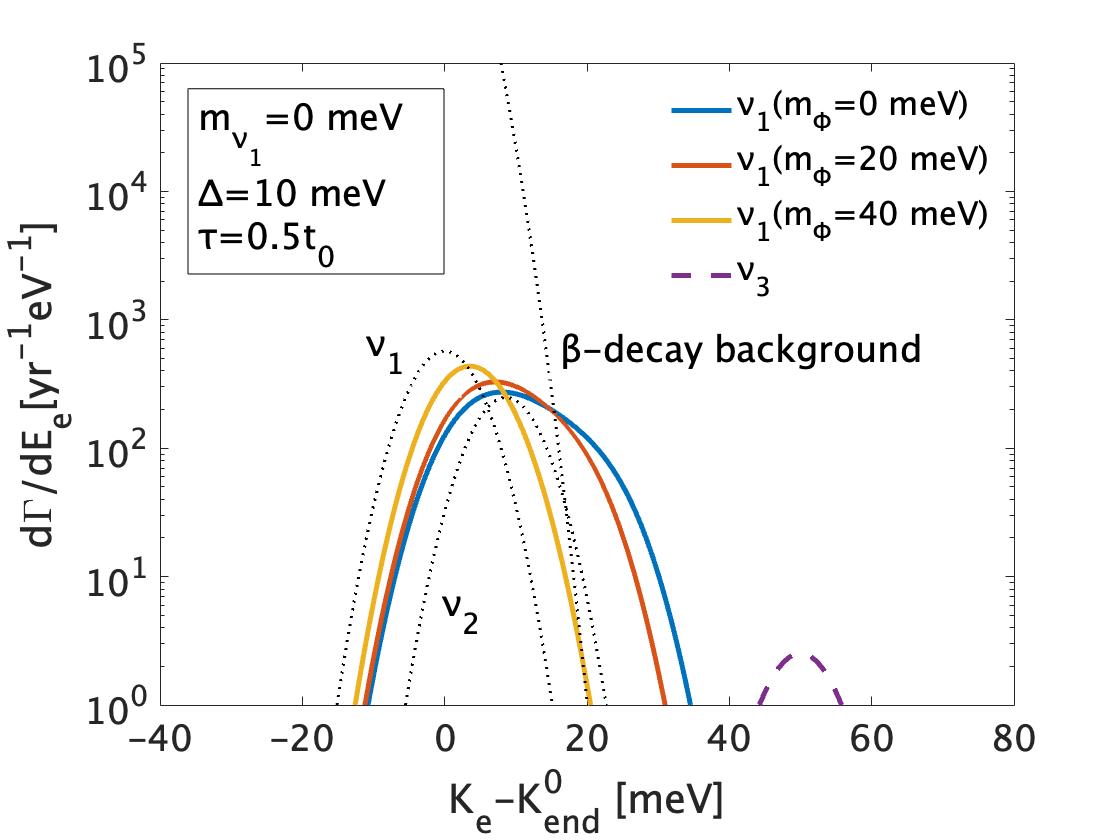}
  \end{center}
 \end{minipage}
  \begin{minipage}{0.5\hsize}
  \begin{center}
   \includegraphics[clip,width=85mm]{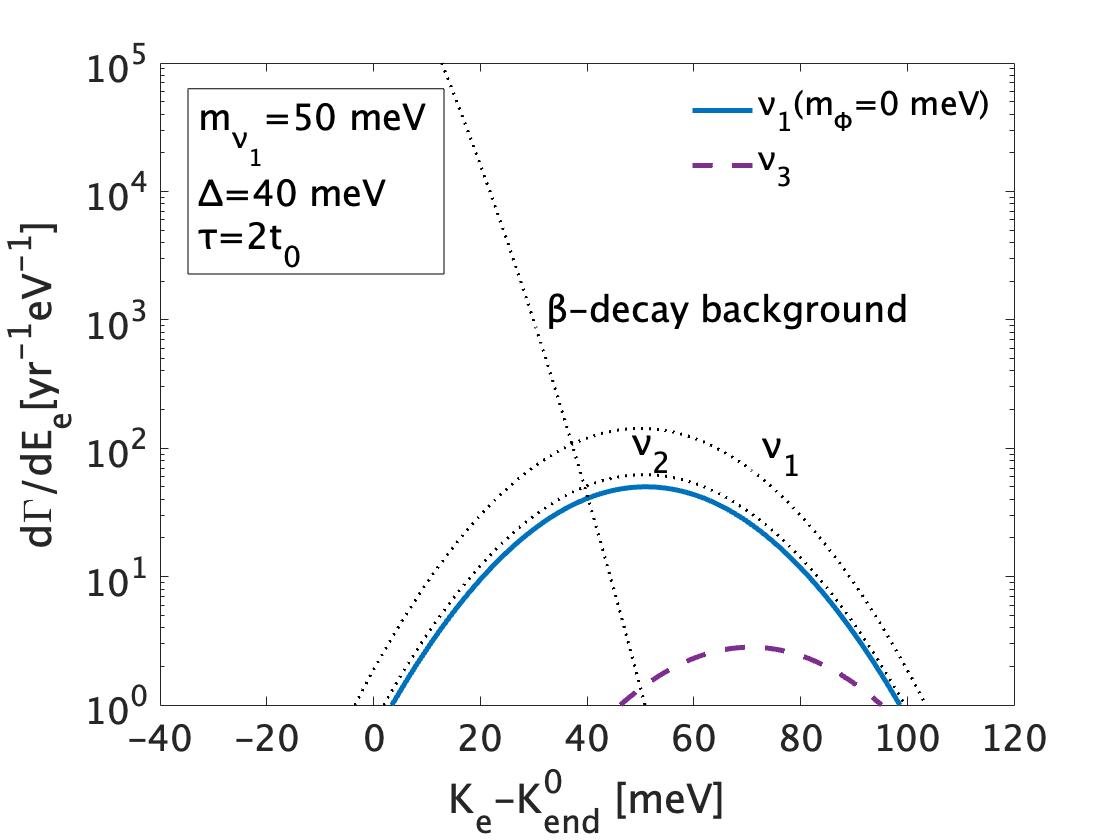}
  \end{center}
 \end{minipage}
  \begin{minipage}{0.5\hsize}
  \begin{center}
   \includegraphics[clip,width=85mm]{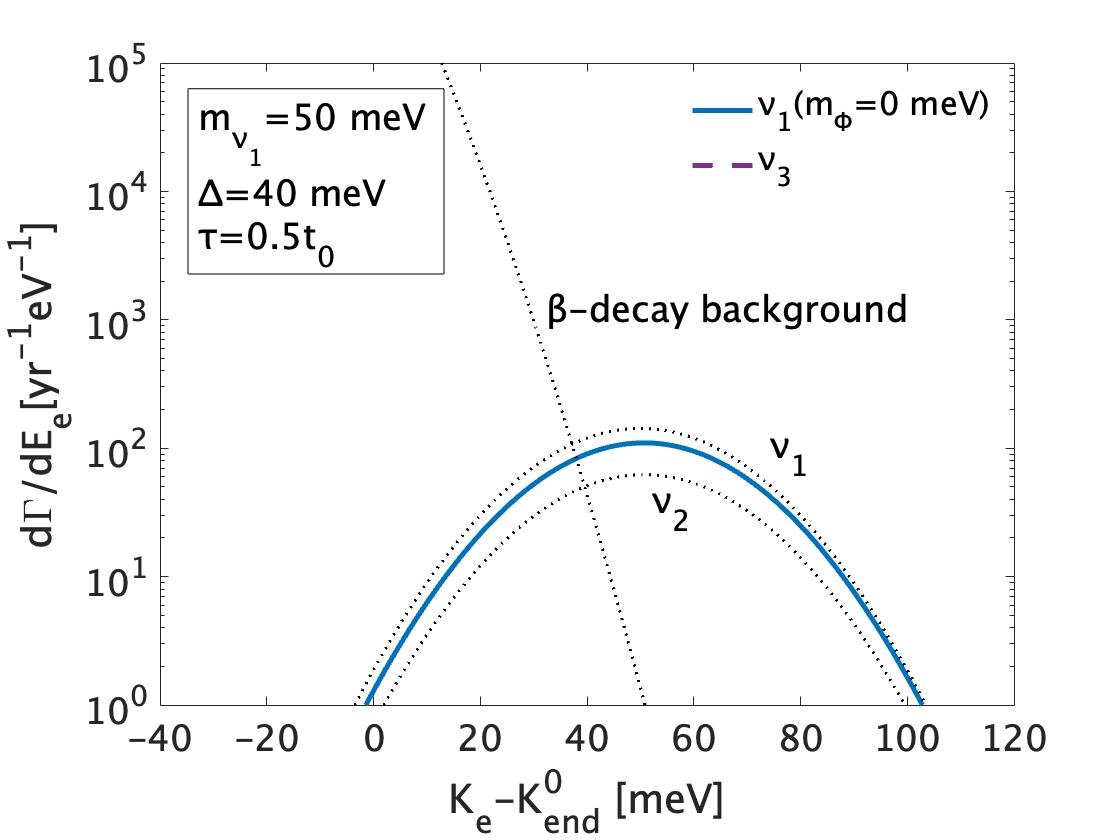}
  \end{center}
 \end{minipage}
  \vspace{-4mm}
 \caption{\small{The expected spectra from the 2-body decays, $\nu_3 \rightarrow \nu_1\phi$, in the neutrino capture on $100$ g of tritium for the NO case. We set $(m_{\nu_1}, \Delta, \tau_{\nu_3})=(0\ {\rm meV},\ 10\ {\rm meV},\ 2t_0)$ (top left panel), $(0\ {\rm meV},\ 10\ {\rm meV},\ 0.5t_0)$ (top right panel), $(50\ {\rm meV},\ 40\ {\rm meV}, 2t_0)$ (bottom left panel) and $(50\ {\rm meV},\ 40\ {\rm meV}, 0.5t_0)$ (bottom right panel) with several values of $m_{\phi}$. Blue, red and yellow lines denote the spectra for $\nu_1$ produced by the decays of $\nu_3$ for $m_{\phi}=0,\ 20,\ 40\ {\rm meV}$, respectively. A purple dashed line denotes the spectrum for $\nu_3$ produced in the early universe but suppressed by the decays of $\nu_3$. Black dotted lines denote the contributions of the spectra for $\nu_1$ and $\nu_2$ produced in the early universe and $\beta$-decay background. Note that the actual spectrum is a superposition of these spectra, and only $m_{\phi}\lesssim 20\ {\rm meV}$ is kinematically allowed for $m_{\nu_1}=50\ {\rm meV}$.}}
 \label{fig:Unstable2_phi}
\end{figure}

First, we discuss the NO case. In Fig.~\ref{fig:Unstable2_phi}, we show the expected spectra in four cases: $(m_{\nu_1}, \Delta)=(0\ {\rm meV},\ 10\ {\rm meV})$ (top panels) and $(50\ {\rm meV},\ 40\ {\rm meV})$ (bottom panels). In the horizontal axis, $K_e \equiv E_e - m_e$ represents the kinetic energy of the emitted electron. In both cases, we consider neutrino lifetimes, $\tau_{\nu_3}= 2t_0$ (left panels) and  $\tau_{\nu_3}=0.5t_0$ (right panels), with several values of the mass of $\phi$, $m_{\phi}$. $t_0=4.35 \times 10^{17}\ {\rm s}$ is the present age of the universe. Blue, red and yellow lines represent the spectra for $\nu_1$ produced by the decays of $\nu_3$ for $m_{\phi}=0,\ 20,\ 40\ {\rm meV}$, respectively. A purple dashed line represents the spectrum for $\nu_3$ produced in the early universe but suppressed by the decays of $\nu_3$. Black dotted lines represent the contributions of the spectra for $\nu_1$ and $\nu_2$ produced in the early universe and $\beta$-decay background. In these figures, the upper threshold energy of the spectrum for $\nu_1$ produced by the decays of $\nu_3$ is characterized by the maximal energy for $\nu_1$, $E^{\ast}$ in Eq.~(\ref{East}). With a larger value of $m_\phi$, the threshold energy for $\nu_1$ produced by the decays of $\nu_3$ becomes smaller since $E^{\ast}$ also becomes smaller.
The lower threshold energy of the spectrum for $\nu_1$ produced by the decays of $\nu_3$ is characterized by the lighter neutrino mass $m_{\nu_1}$ through Eq.~(\ref{relationEeEnu}) with $E_{\nu_1} = m_{\nu_1}$. With a larger value of $m_{\nu_1}$, the lower threshold energy of the spectrum for $\nu_1$ becomes larger although the lower limit of the spectrum for $\nu_1$ produced by the decays of $\nu_3$ cannot be observed experimentally in the case of Fig~\ref{fig:Unstable2_phi}.
Note that due to the energy conservation, $m_{\phi}\lesssim 20\ {\rm meV}$ is allowed for $m_{\nu_1}=50\ {\rm meV}$.

\begin{figure}[htbp]
  \begin{center}
   \includegraphics[clip,width=85mm]{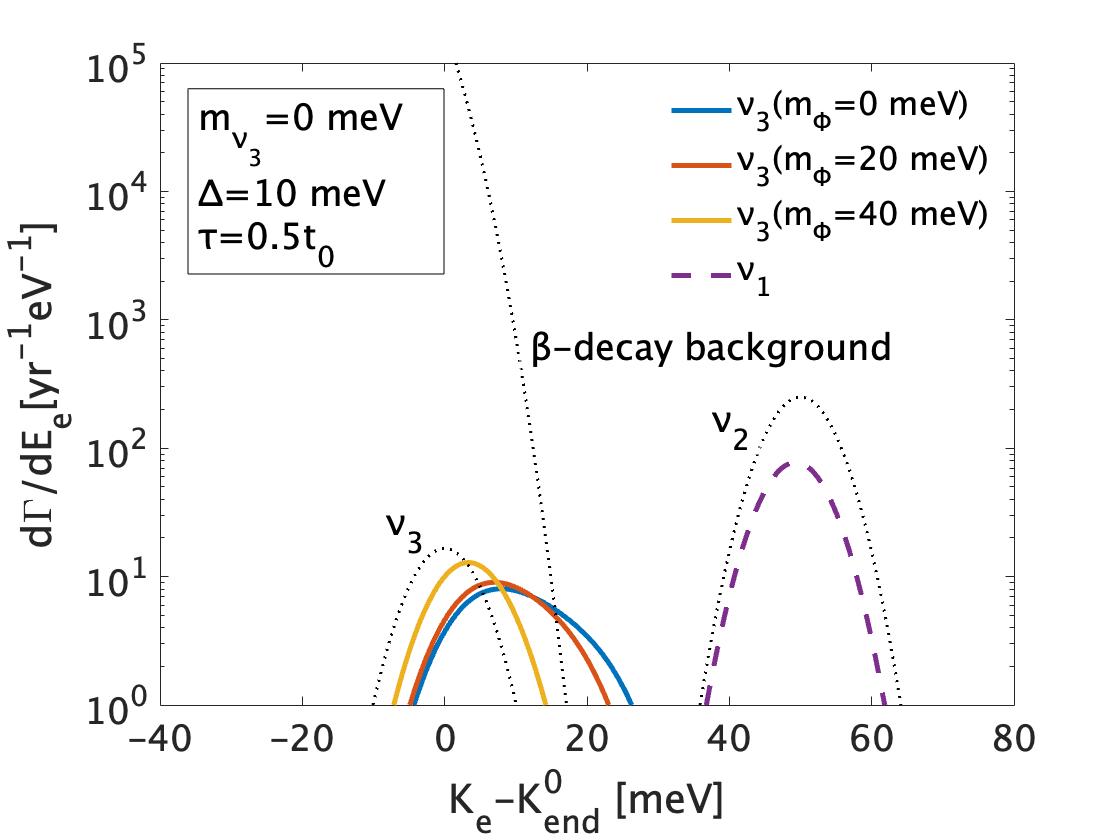}
  \end{center}
  \vspace{-6mm}
 \caption{\small{The expected spectra with $m_{\nu_3}=0\ {\rm meV}$ from the 2-body decays , $\nu_1 \rightarrow \nu_3\phi$, in the neutrino capture on $100$ g of tritium for the IO case. We set $\tau_{\nu_1}=0.5t_0$ with $\Delta=10\ {\rm meV}$ and several values of $m_{\phi}$. Blue, red and yellow lines denote the spectra for $\nu_3$ produced by the decays of $\nu_1$ for $m_{\phi}=0,\ 20,\ 40\ {\rm meV}$, respectively. A purple dashed line denotes the spectrum for $\nu_1$ produced in the early universe but suppressed by the decays of $\nu_1$. Black dotted lines denote the contributions of the spectra of $\nu_2$ and $\nu_3$ produced in the early universe and $\beta$-decay background. Note that the actual spectrum is a superposition of these spectra.}}
 \label{fig:Unstable2_Inverse}
\end{figure}

Finally, we consider the IO case. In Fig.~\ref{fig:Unstable2_Inverse}, we show the expected spectra with $m_{\nu_3}=0\ {\rm meV}$ in the case of the energy resolution of $\Delta= 10\ {\rm meV}$. We set $\tau_{\nu_3}=0.5t_0$ with several values of $m_{\phi}$. In this case, the number of the C$\nu$B signal produced in the early universe increases while the number of the C$\nu$B signal produced by the decays decreases due to the values of the PMNS matrix. 

%%%%%%%%%%%%%%%%%%%%%%%%%%%%%%%%%%%%%%%%%%%%%%%%%%%%%%%%%%%%%%%%%%%%%%%%%%%%%%%%%%%%%%%%%%%%%%%%%%
\subsubsection{3-body decays}

In this section, we consider the expected spectra in the case of 3-body decays.
Hereafter we discuss only the decay channels of $\nu_3 \rightarrow \nu_1 \nu_4\bar{\nu_4}$ in the normal mass ordering and $\nu_1 \rightarrow \nu_3 \nu_4\bar{\nu_4}$ in the inverted mass ordering. As in the previous section, we also set the neutrino masses to satisfy the observed values of neutrino squared-mass differences and the same values of PMNS matrix elements in the case of 2-body decays.

\begin{figure}[htbp]
 \begin{minipage}{0.5\hsize}
  \begin{center}
   \includegraphics[width=85mm]{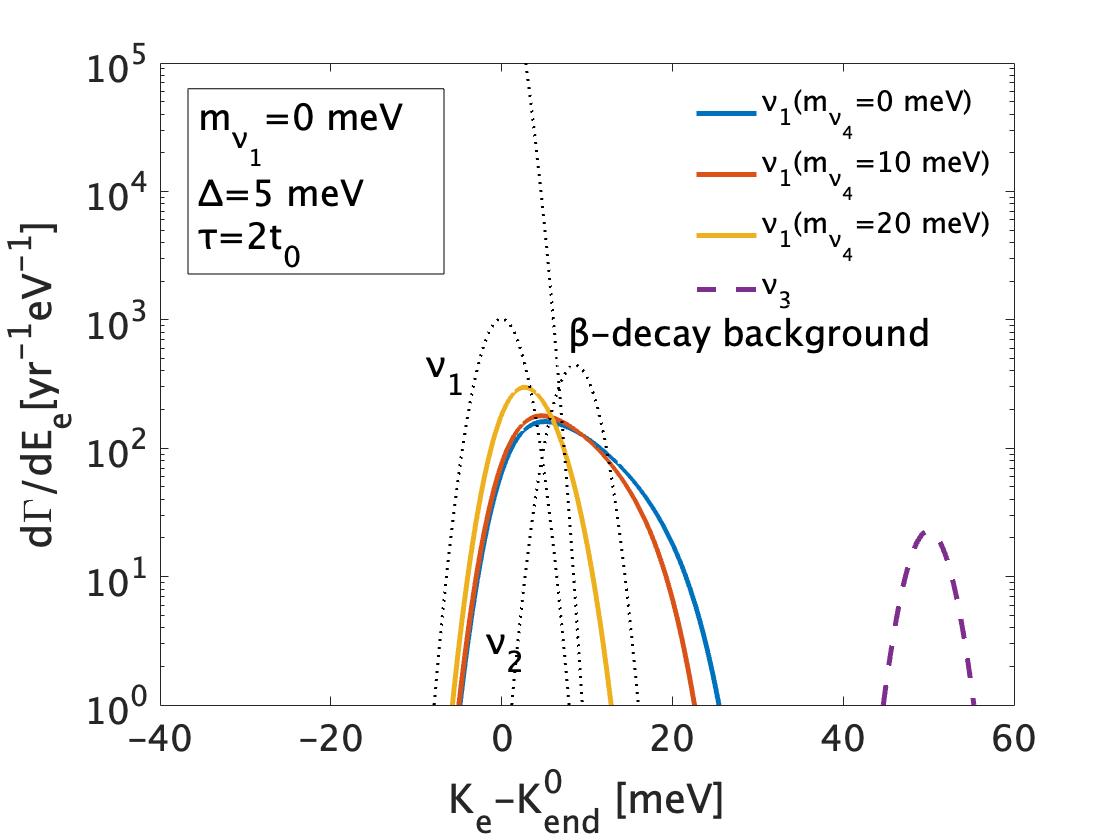}
  \end{center}
 \end{minipage}
 \begin{minipage}{0.5\hsize}
  \begin{center}
   \includegraphics[width=85mm]{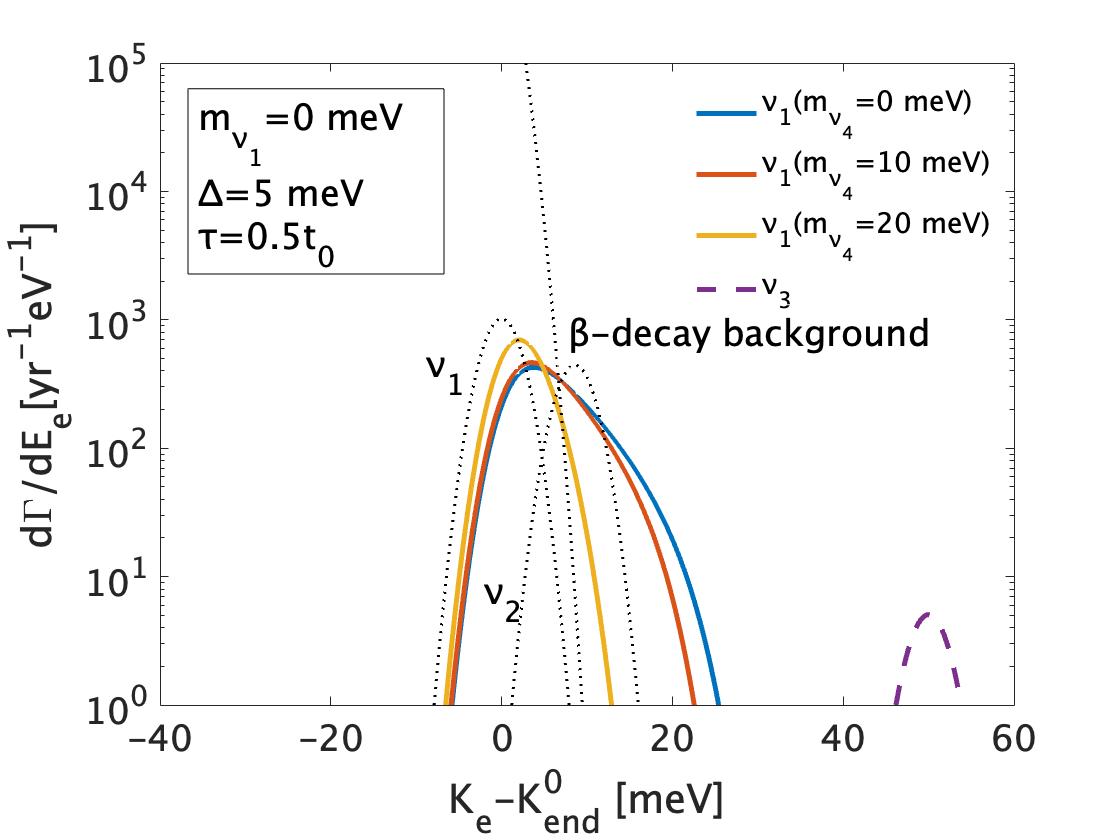}
  \end{center}
 \end{minipage}
 \begin{minipage}{0.5\hsize}
  \begin{center}
   \includegraphics[width=85mm]{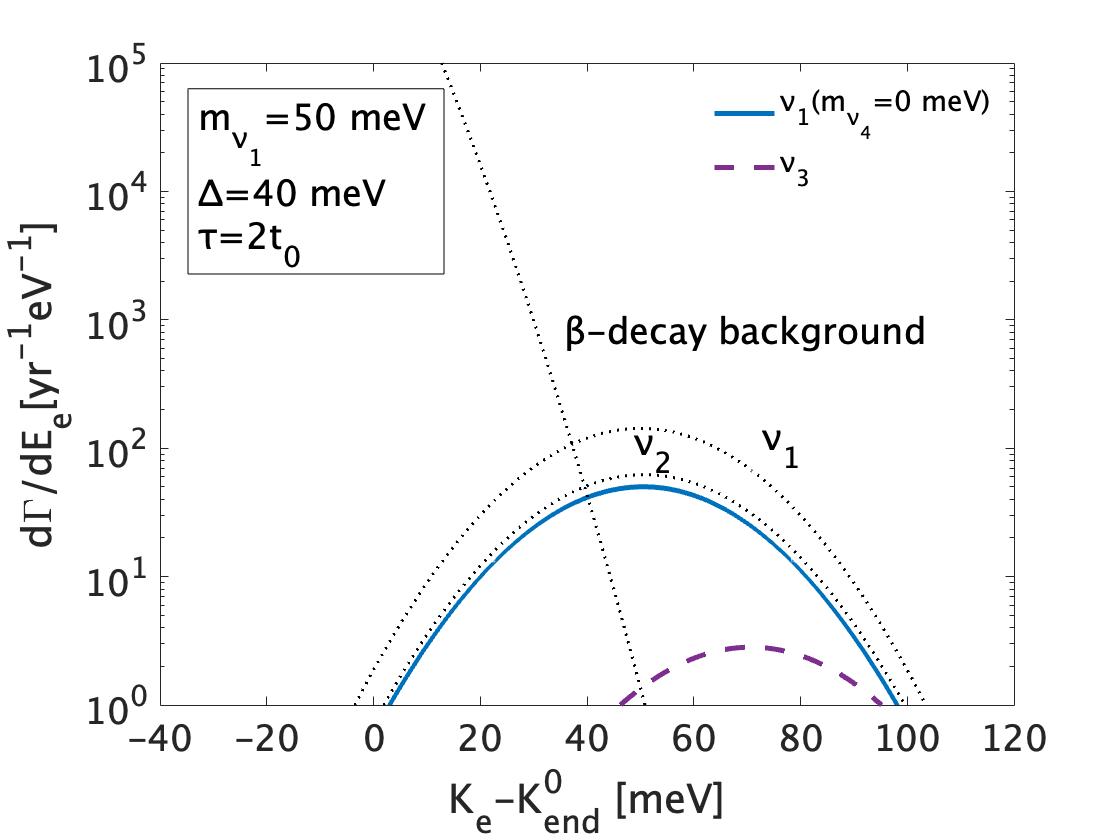}
  \end{center}
 \end{minipage}
 \begin{minipage}{0.5\hsize}
  \begin{center}
   \includegraphics[width=85mm]{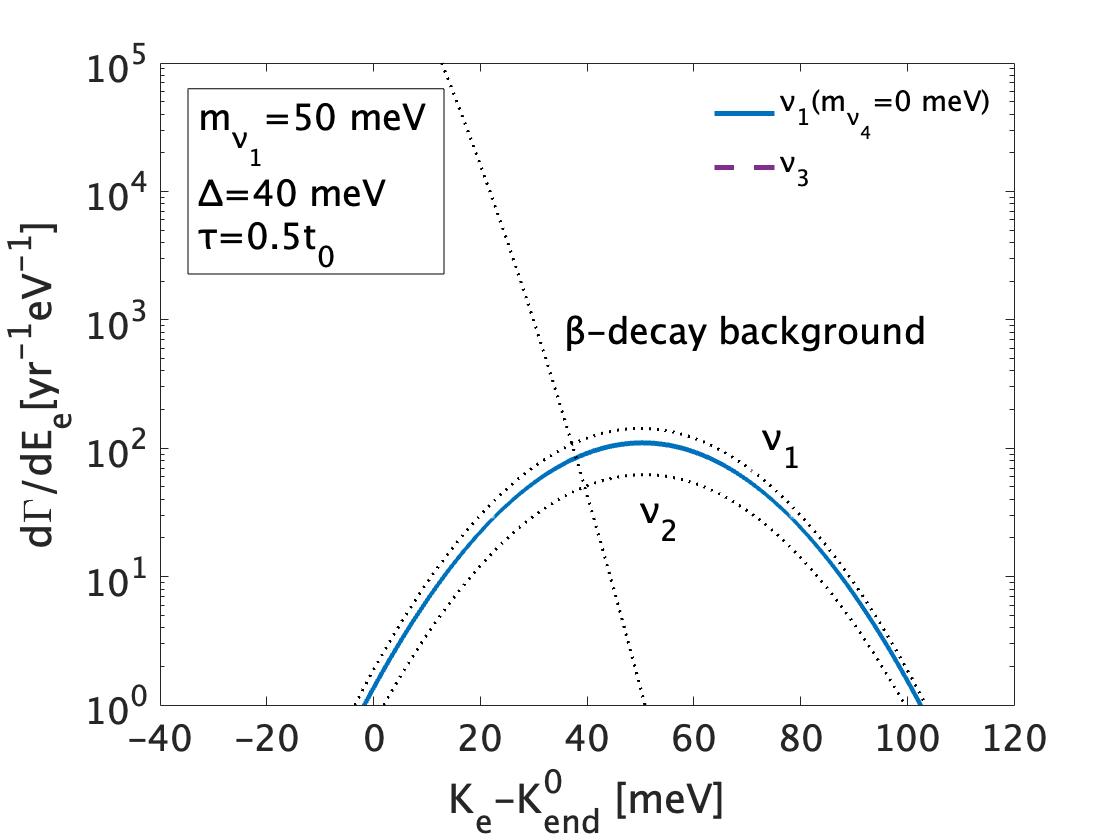}
  \end{center}
 \end{minipage}
  \vspace{-4mm}
 \caption{\small{The expected spectra from the 3-body decays, $\nu_3\rightarrow \nu_1\nu_4\bar{\nu}_4$, in the neutrino capture on $100$ g of tritium for the NO case. We set $(m_{\nu_1}, \Delta, \tau_{\nu_3})=(0\ {\rm meV},\ 5\ {\rm meV},\ 2t_0)$ (top left panel), $(0\ {\rm meV},\ 5\ {\rm meV},\ 0.5t_0)$ (top right panel), $(50\ {\rm meV},\ 40\ {\rm meV}, 2t_0)$ (bottom left panel) and $(50\ {\rm meV},\ 40\ {\rm meV}, 0.5t_0)$ (bottom right panel) with several values of $m_{\phi}$. Blue, red and yellow lines denote the spectra for $\nu_1$ produced by the decays of $\nu_3$ for $m_{\nu_4}=0,\ 10,\ 20\ {\rm meV}$, respectively. A purple dashed line denotes the spectrum for $\nu_3$ produced in the early universe but suppressed by the decays of $\nu_3$. Black dotted lines denote the contributions of the spectra for $\nu_1$ and $\nu_2$ produced in the early universe and $\beta$-decay background. Note that the actual spectrum is a superposition of these spectra.}}
 \label{fig:Unstable3_nu0}
\end{figure}

First, we discuss the NO case. In Fig.~\ref{fig:Unstable3_nu0}, we show the expected spectra in four cases: $(m_{\nu_1}, \Delta)=(0\ {\rm meV},\ 5\ {\rm meV})$ (top panels) and $(50\ {\rm meV},\ 40\ {\rm meV})$ (bottom panels). In both cases, we also consider neutrino lifetimes, $\tau_{\nu_3}= 2t_0$ (left panels) and  $\tau_{\nu_3}=0.5t_0$ (right panels), with several values of the mass of $\phi$, $m_{\phi}$. Blue, red and yellow lines represent the spectra for $\nu_1$ produced by the decays of $\nu_3$ for $m_{\nu_4}=0,\ 10,\ 20\ {\rm meV}$, respectively. A purple dashed line represents the spectrum for $\nu_3$ produced in the early universe but suppressed by the decays of $\nu_3$. Black dotted lines represent the spectra for $\nu_1$ and $\nu_2$ produced in the early universe, and $\beta$-decay background. In these figures, the upper threshold energy of the spectrum for $\nu_1$ produced by the decays of $\nu_3$ is characterized by $E_{\nu_1}^{\rm max}$ in Eq.~(\ref{Emax3}). Thus, a larger value of $m_{\nu_4}$ reduces the upper limit of energy for $\nu_1$. As in the case of 2-body decays, a larger value of $m_{\nu_1}$ increases the lower threshold energy. Note that due to energy conservation, only $m_{\nu_4}\lesssim 25\,(10) \ {\rm meV}$ is allowed for $m_{\nu_1} = 0\,(50)\ {\rm meV}$.

The lower panels of Fig.~\ref{fig:Unstable3_nu0} for $m_{\nu_1}=50\ {\rm meV}$ are almost the same with the case of 2-body decays for $m_{\nu_1}=50\ {\rm meV}$ in the lower panels of Fig.~\ref{fig:Unstable2_phi} since in both cases of 2-body and 3-body decays, $\nu_1$ produced by the decays of $\nu_3$ are non-relativistic in the current universe. In these cases, we cannot distinguish 2-body and 3-body decays without an extremely small energy resolution. 

\begin{figure}[h]
  \begin{center}
   \includegraphics[clip,width=85mm]{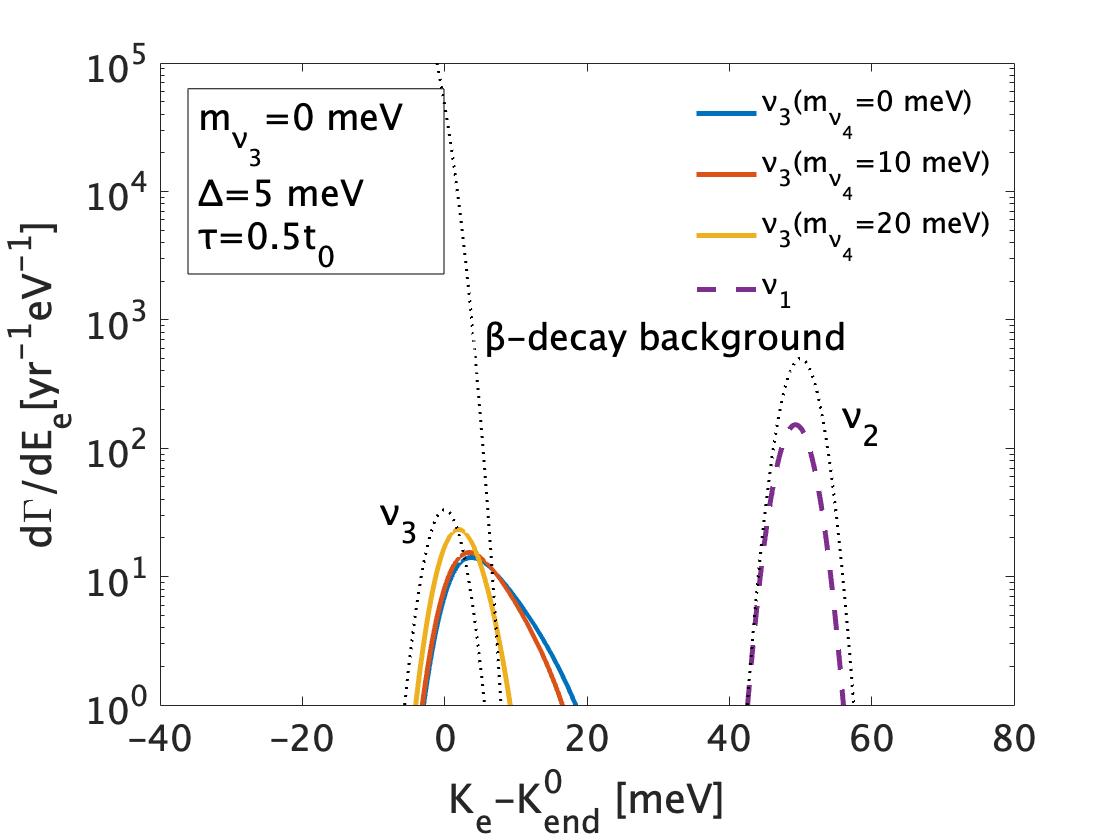}
  \end{center}
  \vspace{-6mm}
 \caption{\small{The expected spectra with $m_{\nu_3}=0\ {\rm meV}$ from the 3-body decays, $\nu_1\rightarrow \nu_3\nu_4\bar{\nu}_4$, in the neutrino capture on $100$ g of tritium for the IO case. We set $\tau_{\nu_1}=0.5t_0$ with $\Delta=5\ {\rm meV}$. and several values of $m_{\nu_4}$. Blue, red and yellow lines denote the spectra for $\nu_3$ produced by the decays of $\nu_1$ for $m_{\nu_4}=0,\ 10,\ 20\ {\rm meV}$, respectively. A purple dashed line denotes the spectrum for $\nu_1$ produced in the early universe but suppressed by the decays of $\nu_1$. Black dashed lines denote the contributions of the spectra for $\nu_2$ and $\nu_3$ produced in the early universe and $\beta$-decay background. Note that the actual spectrum is a superposition of these spectra.}}
 \label{fig:Unstable3_Inverse}
\end{figure}

Finally, we show results for the IO case. In Fig.~\ref{fig:Unstable3_Inverse}, we show the expected spectra with $m_{\nu_3}=0\ {\rm meV}$ in the case of the energy resolution of $\Delta= 5\ {\rm meV}$. We set $\tau_{\nu_3}= 0.5t_0$ with several values of $m_{\nu_4}$.

\clearpage
%%%%%%%%%%%%%%%%%%%%%%%%%%%%%%%%%%%%%%%%%%%%%%%%%%%%%%%%%%%%%%%%%%%%%%%%%%%%%%%%%%%%%%%%%%%%%%%%%%
%%%%%%%%%%%%%%%%%%%%%%%%%%%%%%%%%%%%%%%%%%%%%%%%%%%%%%%%%%%%%%%%%%%%%%%%%%%%%%%%%%%%%%%%%%%%%%%%%%
%%%%%%%%%%%%%%%%%%%%%%%%%%%%%%%%%%%%%%%%%%%%%%%%%%%%%%%%%%%%%%%%%%%%%%%%%%%%%%%%%%%%%%%%%%%%%%%%%%
%%%%%%%%%%%%%%%%%%%%%%%%%%%%%%%%%%%%%%%%%%%%%%%%%%%%%%%%%%%%%%%%%%%%%%%%%%%%%%%%%%%%%%%%%%%%%%%%%%
\subsection{Constraints on a neutrino lifetime and an invisible particle mass}
\label{Sec5.2}

\subsubsection{Method}
In order to forecast how well the PTOLEMY-type experiment will be able to constrain a neutrino lifetime and the mass of invisible particles, we employ a $\chi^2$ test to the would-be observed C$\nu$B signals.

To construct a Gaussian $\chi^2$-function, we refer to the formulation proposed in the PTOLEMY project \cite{Betti:2019ouf} (see also \cite{KATRIN:2005fny, Nucciotti:2009wq, SejersenRiis:2011sj}). We define the would-be observed numbers of $\beta$-decay background and the C$\nu$B signals within an energy bin $i$ with width $\Delta$, which is the same with the energy resolution of the detector, centered at $E_i$,
\begin{align}
&N_{\beta}^i={\mathrm T}\int^{E_i+\Delta/2}_{E_i-\Delta/2}\frac{d\bm{{\Gamma}}_{\beta}}{dE_e}dE_e, \nonumber \\
&N_{\rm C\nu B}^i={\mathrm T}\int^{E_i+\Delta/2}_{E_i-\Delta/2}\frac{d\bm{{\Gamma}}_{\rm C\nu B}}{dE_e}dE_e, \nonumber \\
&\tilde{N}_{\rm C\nu B}^i={\mathrm T}\int^{E_i+\Delta/2}_{E_i-\Delta/2}\frac{d\tilde{\bm{{\Gamma}}}_{\rm C\nu B}}{dE_e}dE_e,
\end{align}
where ${\mathrm T}$ is the exposure time. For fiducial models, the total number of expected observable signals per an energy bin is given by
\begin{align}
N_{\rm obs}^i(\hat{\bm{\theta}})= N_{\beta}^i(\hat{\bm{\theta}})+N_{\rm C\nu B}^i(\hat{\bm{\theta}})+\tilde{N}_{\rm C\nu B}^i(\hat{\bm{\theta}}),
\end{align}
where $\bm{\theta}$ is the vector of parameters and $\hat{\bm{\theta}}$ is that for the fiducial values. 
In this formulation, the number of signals per an energy bin for a parameter $\bm{\theta}$ is also given by
\begin{align}
N^i(\bm{\theta})=N_{\beta}^i(\bm{\theta})+N_{\rm C\nu B}^i(\bm{\theta})+\tilde{N}_{\rm C\nu B}^i(\bm{\theta}).
\end{align} 
Then we can obtain a Gaussian $\chi^2$-function, assuming a statistical error of $\sqrt{N_{\rm obs}^i(\hat{\bm{\theta}})}$ in each bin,
\begin{align}
\chi^2(\bm{\theta})=\sum_i\left(\frac{N^i(\bm{\theta})-N_{\rm obs}^i(\hat{\bm{\theta}})}{\sqrt{N^i_{\rm obs}(\hat{\bm{\theta}})}} \right)^2,
\label{chi2}
\end{align}
which is translated into the corresponding likelihood function by $\chi^2=-2\log \mathcal{L}$.
\footnote{For energy bins with low statistics, a Gaussian $\chi^2$-function may not be appropriate, strictly speaking. However, we confirmed that the following constraints on a neutrino lifetime and the mass of invisible particles change by a factor but not an order even if we consider a Poissonian likelihood instead of a Gaussian likelihood,
\begin{align}
    \mathcal{L}(\bm{\theta})=\prod_{i}\frac{N^i(\bm{\theta})^{N^i_{\rm obs}(\hat{\bm{\theta}})}e^{-N^i(\bm{\theta})}}{\Gamma[N^i_{\rm obs}(\hat{\bm{\theta}})+1]}.
    \label{Plikelihood}
\end{align}
In particular, the constraints on the mass of invisible particles are slightly improved in the case of a Poissonian likelihood. A detailed comparison between Gaussian and Poissonian likelihoods will be important when the setup of the PTOLEMY-type experiment is more precisely determined, but here we leave it for a future issue and take a simple Gaussian likelihood.} 

For our analysis, we assume an observed energy range of $E_i$ between $E_{\rm min}=E_{\rm end}^0$ and $E_{\rm max}=E_{\rm end}^0+100, 150, 300, 600\ {\rm meV}$ for $m_{\rm lightest}=0, 50, 100, 200\ {\rm meV}$, respectively, where $E_{\rm end}^0\equiv K_{\rm end}^0+m_e$ is the endpoint energy for massless neutrinos. 
We set $E_{\rm max}$ to be a sufficiently high energy in the no-signal region whereas $E_{\rm min}$ is determined by the smallest value of the energy for signal from the neutrino decays. 
When the value of $E_{\rm min}$ changes from $E_{\rm end}^0-\Delta/2$ to $E_{\rm end}^0+\Delta/2$, the constraints on a neutrino lifetime and the invisible particle mass slightly change by a factor since the intervals of the energy bins shift.
The detailed value for $E_{\rm min}$ will be determined when the setup is more concrete and the observational data is actually obtained.
We also take several values of the energy resolution, $\Delta$, and an exposure, ${\rm TM_T}$, where ${\rm M_T}$ is the total tritium mass.

With some fiducial values of parameters $\hat{\bm{\theta}}$ fixed, we compute $\Delta\chi^2\equiv \chi^2-\chi_{\rm min}^2$, where $\chi^2_{\rm min}$=0 for the fiducial values in the definition of Eq.~(\ref{chi2}), on a grid of $\bm{\theta}=(\tau_{\nu_i},\ m_{\phi})$ for 2-body decays, $\nu_i \rightarrow \nu_j\phi$, and  $\bm{\theta}=(\tau_{\nu_i},\ m_{\nu_4})$ for 3-body decays, $\nu_i \rightarrow \nu_j\nu_4\bar{\nu}_4$.
For $\tau_{\nu_i}$, we take 400 grid points as $\tau_{\nu_i}^{k,l}=10^k+(l-1)[10^{k+1}-10^{k}]/99\ (k=-2,-1,0,1,\ l=1,...100)$ in the region $\tau_{\nu_i}^{k,l}\in [10^{-2},\ 10^2]$.
For $m_{\phi}$ and $m_{\nu_4}$, we take a width of grid, $\Delta_m=0.5\ {\rm meV}$ in the region $m_{\phi},\ m_{\nu_4}\in[0,\ m_{\rm max}]$, where $m_{\rm max}$ is the kinematically allowed maximal value of $m_{\phi}$ or $m_{\nu_4}$.

Subsequently $\chi^2$ is given marginalized over $m_{\phi}$ or $m_{\nu_4}$ in some cases. 
Marginalized 1-D $\chi^2$, $\chi^2_{\rm 1-D}$ , is defined by
\begin{align}
\chi^2_{\rm 1-D}(\tau_{\nu_i})&=-2\log \mathcal{L}_{\rm 1-D}, \nonumber \\
\mathcal{L}_{\rm 1-D}&=\int dm\  \exp\left[-\frac{1}{2}\chi^2(\bm{\theta}) \right],
\label{Mchi2}
\end{align}
where $m=m_{\phi}$ for 2-body decays, $\nu_i \rightarrow \nu_j\phi$, $m=m_{\nu_4}$ for 3-body decays, $\nu_i \rightarrow \nu_j\nu_4\bar{\nu}_4$, and we assume the prior probability is flat.
Then we estimate $\Delta \chi^2_{\rm 1-D}\equiv \chi^2_{\rm 1-D}-\chi^2_{\rm 1-D,\ min}$, where $\chi^2_{\rm 1-D,\ min}=\underset{\tau_{\nu_i}}{\min}\left[\chi^2_{\rm 1-D}(\tau_{\nu_i})\right]$.
Considering the marginalized 1-D $\chi^2$, we discuss a parameter to be ruled out at $1\sigma$ when $\sqrt{\Delta\chi^2_{\rm 1-D}}\geq1$ and/or at $n\sigma$ when $\sqrt{\Delta\chi^2_{\rm 1-D}}\geq n$. 

Here we comment on how we handle the masses and the ordering of active neutrinos in our analysis. Currently, the mass of the lightest neutrinos, $m_{\rm lightest}$, and the ordering of neutrino masses are not determined since we only know the two squared-mass differences from neutrino oscillation experiments. In the future PTOLEMY-type experiment with 100 g of tritium, we will determine the value of $m_{\rm lightest}$ and their ordering by observing a large number of events of $\beta$-decay background (see Figs.~2 and 3 of ref.~ \cite{Betti:2019ouf}) since the shape of the $\beta$-decay spectrum near the endpoint significantly depends on $m_{\rm lightest}$ and neutrino mass ordering. Although the $\beta$-decay spectrum at the endpoint energy $E_{\rm end}$ overlaps with the C$\nu$B signal produced both by the decay and during the early universe, even in the region of $E_e< E_{\rm end}$, the $\beta$-decay spectrum is still dependent on $m_{\rm lightest}$ and in this region the $\beta$-decay spectrum is the dominated signal. Even if $m_{\rm lightest}$ is slightly deviated from the fiducial value $\hat{m}_{\rm lightest}$, the contributions of the $\beta$-decay spectrum to $\chi^2$ in the region of $E_e< E_{\rm end}$ are very large and $m_{\rm lightest}$ would be severely constrained.

Therefore, in our analysis, we neither marginalize $m_{\rm lightest}$ nor the mass ordering, rather fix them as the would-be known values.
In the case of 2-body decays with $(\hat{\tau}_{\nu_i}, \hat{m}_{\phi,\ \nu_4})=(t_0,\ 0)$, we confirmed that we obtain the same result after maginalizing $m_{\rm lightest}$ with different $10\%$ and $50\%$ uncertainties for a fiducial value of the lightest neutrino mass $\hat{m}_{\rm lighest}$ by using 10 grid points equally spaced.
For $\hat{m}_{\rm lightest}=0$, we also confirmed that we obtain the same result by taking 10 grind points in the region $m_{\rm lightest}\in [0,\ 10\ {\rm meV}]$.
(Note that in these cases we set $E_{\rm min}=E^0_{\rm end}-100,150,300,600\ {\rm meV}$ for $\hat{m}_{\rm lightest}=0, 50, 100, 200\ {\rm meV}$, respectively, to measure $\beta$-decay background well.)

%%%%%%%%%%%%%%%%%%%%%%%%%%%%%%%%%%%%%%%%%%%%%%%%%%%%%%%%%%%%%%%%%%%%%%%%%%%%%%%%%%%%%%%%%%%%%%%%%%

\subsubsection{2-body decays: $\nu_{3} \rightarrow \nu_4\phi$ in the NO case, $\nu_{1} \rightarrow \nu_4\phi$ in the IO case, and any other processes of decays of $\nu_3$ in the NO case and of $\nu_1$ in the IO case}
\label{Sec:5.2.3}

In this section, we forecast the constraints on neutrino lifetimes in the 2-body decays, $\nu_{3} \rightarrow \nu_4\phi$ in the NO case and $\nu_{1} \rightarrow \nu_4\phi$ in the IO case.
In this case, we cannot observe the expected spectrum for $\nu_4$ produced by the decays of $\nu_3$ in the NO case and $\nu_1$ in the IO case\footnote{We assume $U_{\rm e4}=0$ in the PMNS matrix.}. However, we will impose the constraints on neutrino lifetimes through the suppression of the observed C$\nu$B signal. Here the suppressed signal is $\nu_3$ in the NO case and $\nu_1$ in the IO case. Then $\bm{\theta}$ is simplified to be $\theta=\tau_{\nu_{3}}$ in the NO case and $\theta=\tau_{\nu_{1}}$ in the IO case and the result is independent of $m_{\nu_4}$ as long as the process of 2-body decays is kinematically allowed. This forecast can be applied to any processes of decays of $\nu_3$ in the NO case and of $\nu_1$ in the IO case too.

\begin{figure}[h]
  \begin{center}
   \includegraphics[clip,width=85mm]{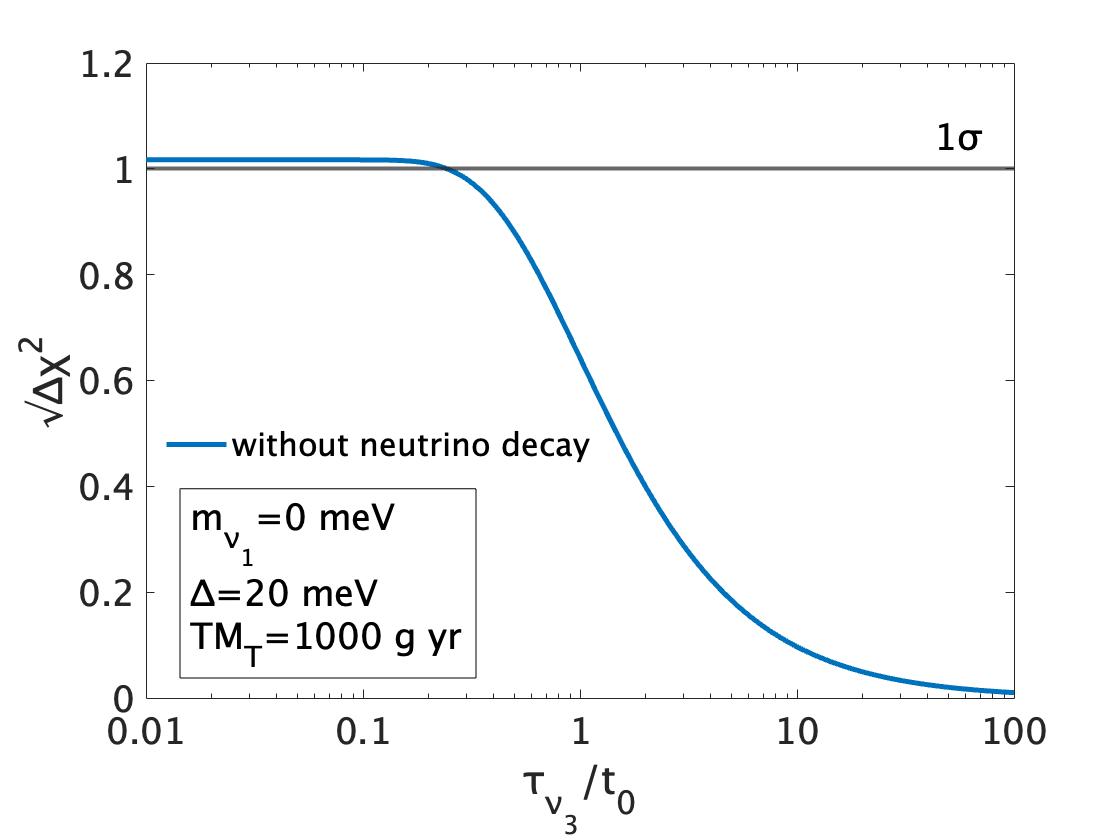}
  \end{center}
  \vspace{-6mm}
 \caption{\small{$\sqrt{\Delta\chi^2}$ as a function of a test value of $\tau_{\nu_3}$ for $\nu_{3} \rightarrow \nu_4\phi$ in the NO case with $m_{\nu_1}=0\ {\rm meV}$, where a fiducial value of $\hat{\tau}_{\nu_3}$ is assumed to be $\infty$, suggested by the SM. An exposure of ${\rm TM_T}=1000$ g yr and an energy resolution of $\Delta=20\ {\rm meV}$ are also assumed.}}
 \label{fig:chi2_decay0_NH}
\end{figure}

First, we consider $\nu_3\rightarrow \nu_4\phi$ in the NO case. Unfortunately, in this case, we cannot observe many events of the signal for the heaviest neutrinos $\nu_3$ because of too small $|U_{e3}|$, and we cannot impose severe constraints on a neutrino lifetime without a large exposure. 
In Fig.~\ref{fig:chi2_decay0_NH}, we show the constraints on the neutrino lifetime $\tau_{\nu_3}$ for a fiducial value of $\tau_{\nu_3}$ with $\hat{\tau}_{\nu_3}=\infty$ (no decay). Here we set a large exposure of $1000$ g yr, $\Delta=20\ {\rm meV}$ and $m_{\nu_1}=0\ {\rm meV}$. The lower bound at $1\sigma$ confidence level is $\tau_{\nu_3}\simeq 0.2t_0$, which is much longer than the current constraints on $\tau_{\nu_3}$. 
Due to the small signal for $\nu_3$, we cannot impose an upper bound for a fiducial value of $\tau_{\nu_3}$ for unstable cosmic neutrinos with $\hat{\tau}_{\nu_3}\neq \infty$. 
If the lightest neutrinos are more massive, we can distinguish the C$\nu$B signal from the $\beta$-decay background with a larger energy resolution since the $\beta$-decay background is far away from the C$\nu$B signal as in figures in section \ref{Sec5.1}. However, in the NO case, it is difficult to distinguish the signal for the massive lightest neutrinos $\nu_1$ with the large $|U_{e1}|$ from the suppressed signal for $\nu_3$ by their decays since these signals are heavily overlapped as in the lower panels of Fig.~\ref{fig:Unstable2_phi}. Then the required energy resolution will be almost the same or worse compared with the case of the massless $\nu_1$ since we cannot distinguish the statistical fluctuation for massive $\nu_1$ from the suppressed signal for $\nu_3$ due to their decays.
 
\begin{figure}[h]
 \begin{minipage}{0.5\hsize}
  \begin{center}
   \includegraphics[width=85mm]{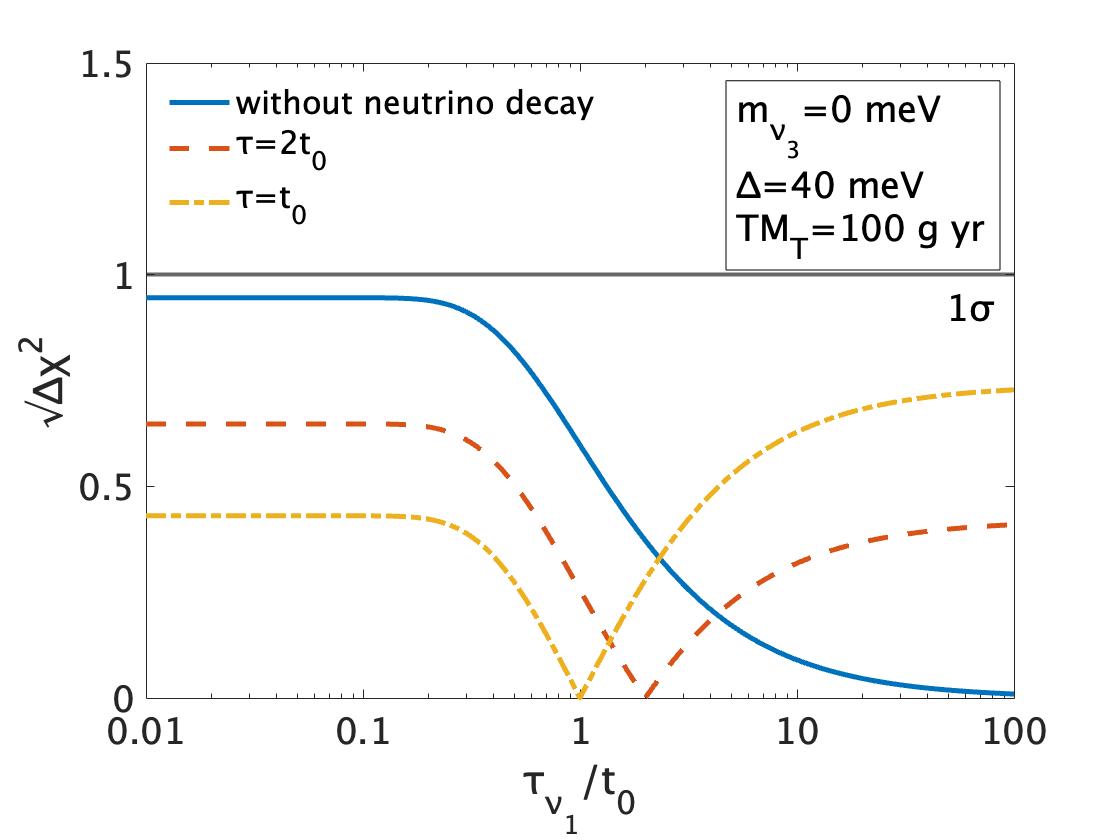}
  \end{center}
 \end{minipage}
 \begin{minipage}{0.5\hsize}
  \begin{center}
   \includegraphics[width=85mm]{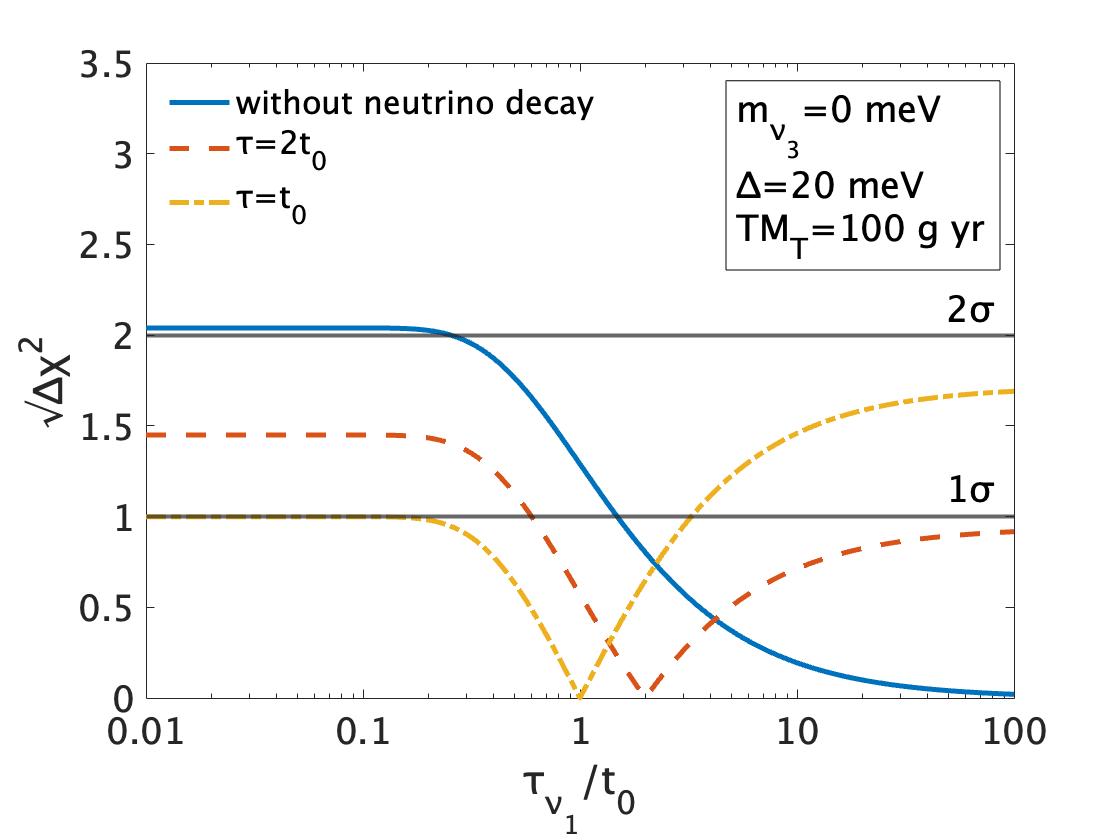}
  \end{center}
 \end{minipage}
 \begin{minipage}{0.5\hsize}
  \begin{center}
   \includegraphics[width=85mm]{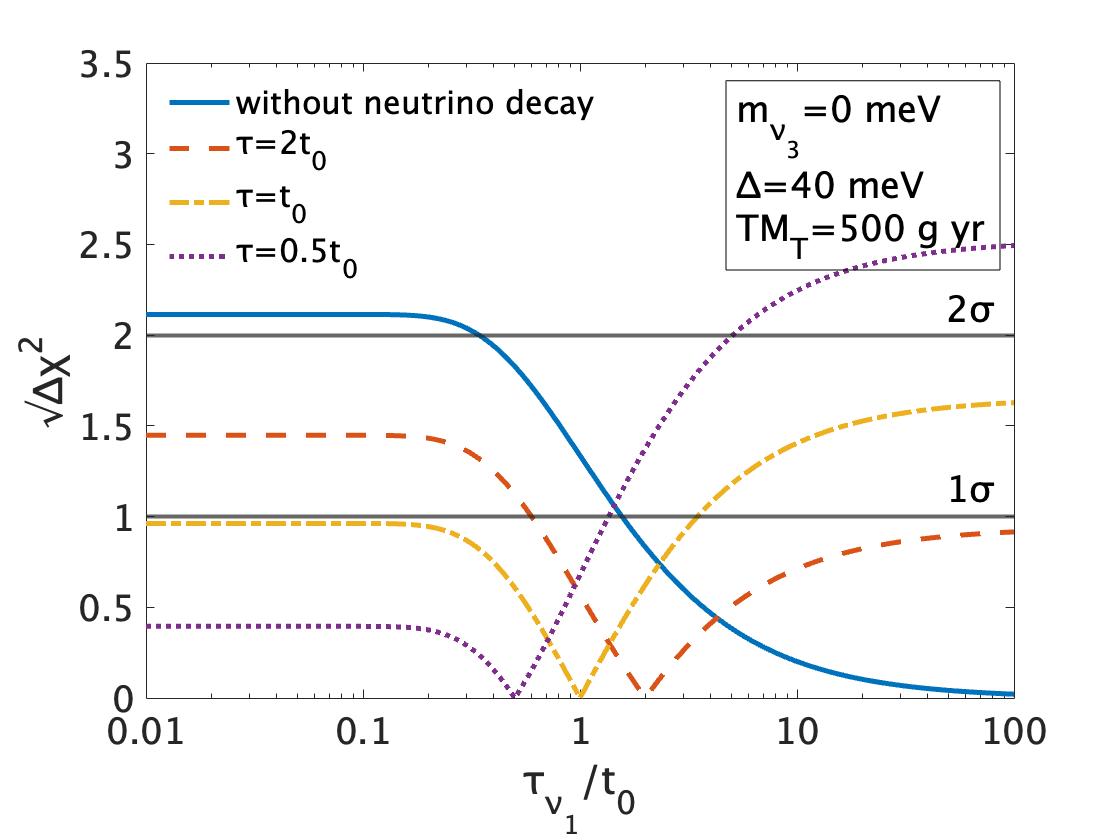}
  \end{center}
 \end{minipage}
 \begin{minipage}{0.5\hsize}
  \begin{center}
   \includegraphics[width=85mm]{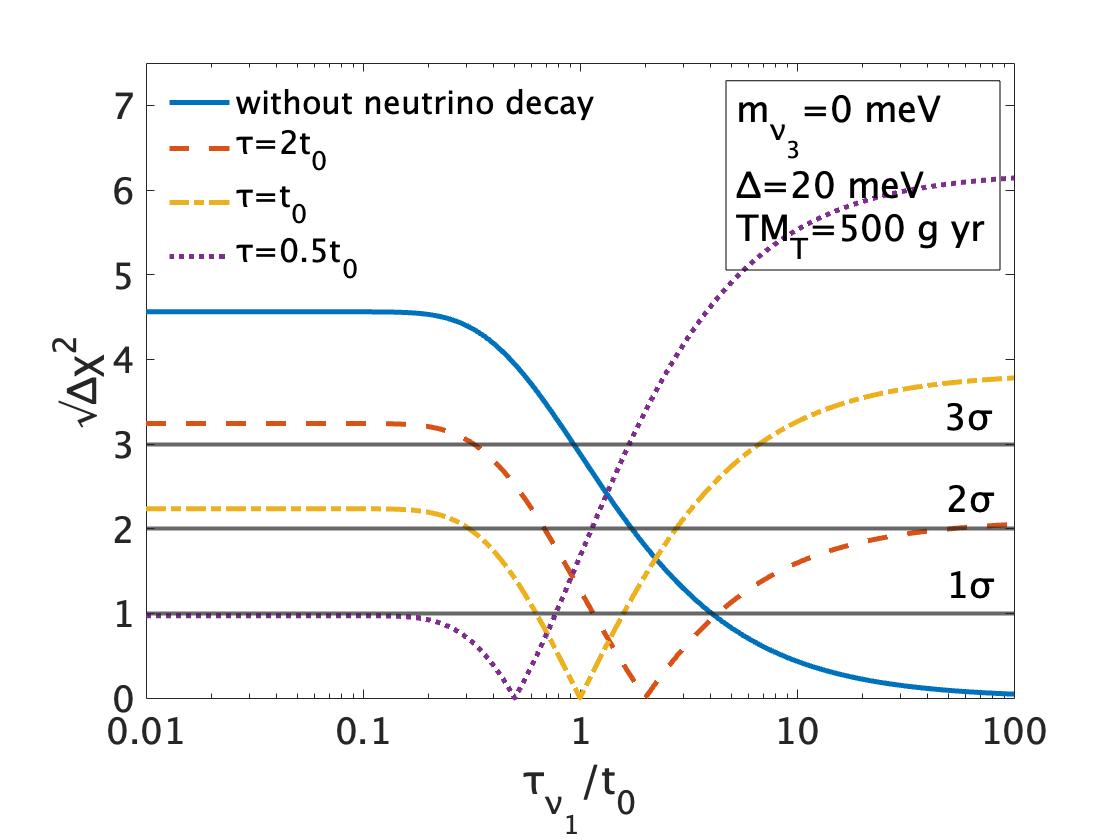}
  \end{center}
 \end{minipage}
  \vspace{-4mm}
 \caption{\small{$\sqrt{\Delta\chi^2}$ as a function of a test value of $\tau_{\nu_1}$ for  $\nu_{1} \rightarrow \nu_4\phi$ in the IO case with $m_{\nu_3}=0\ {\rm meV}$, where fiducial values of $\tau_{\nu_1}$ are assumed to be $\infty$ (blue solid), $2t_0$ (red dashed), $t_0$ (yellow dash-dotted), $0.5t_0$ (purple dotted). We set $({\rm TM_T},\ \Delta)=(100\ {\rm g\ yr},\ 40\ {\rm meV})$ (top left panel), $(100\ {\rm g\ yr},\ 20\ {\rm meV})$ (top right panel), $(500\ {\rm g\ yr},\ 40\ {\rm meV})$ (bottom left panel) and $(500\ {\rm g\ yr},\ 20\ {\rm meV})$ (bottom right panel). In an exposure of 100 g yr, the signal of C$\nu$B with $\tau_{\nu_1}\lesssim0.5t_0$ cannot be observed since the number of this event is smaller than one \cite{Akita:2020jbo}.}}
 \label{fig:chi2_decay0_IH}
\end{figure}

In Fig.~\ref{fig:chi2_decay0_IH}, we show the constraints on the neutrino lifetime $\tau_{\nu_1}$ for $\nu_1\rightarrow \nu_4\phi$ in the IO case for four different fiducial values: $\hat{\tau}_{\nu_1}=\infty$ (blue solid), $2t_0$ (red dashed), $t_0$ (yellow dash-dotted), $0.5t_0$ (purple dotted). 
We assume $m_{\nu_3}=0\ {\rm meV}$, and $({\rm TM_T},\ \Delta)=(100\ {\rm g\ yr},\ 40\ {\rm meV})$ (top left panel), $(100\ {\rm g\ yr},\ 20\ {\rm meV})$ (top right panel), $(500\ {\rm g\ yr},\ 40\ {\rm meV})$ (bottom left panel) and $(500\ {\rm g\ yr},\ 20\ {\rm meV})$ (top left panel). Note that in an exposure of 100 g yr, the signal of $\nu_1$ with $\tau_{\nu_1}=0.5t_0$ cannot be observed since the expected number of this event is smaller than unity \cite{Akita:2020jbo}. 
Due to the large value of $|U_{e1}|$, the constraints on a neutrino lifetime are more severe than those in the NO case.
Even using the detector with its energy resolution of $\Delta=40\ {\rm meV}$ in an exposure of $500$ g yr, we will be able to impose an upper bound of $\tau_{\nu_1}\lesssim 4t_0$ at $2\sigma$ confidence level if a fiducial value of $\tau_{\nu_1}$ is $\hat{\tau}_{\nu_1}\simeq 0.5t_0$, and then we can distinguish this scenario from the case with $\hat{\tau}_{\nu_1}=\infty$ (no decay). In the case of $\Delta\lesssim20 {\rm meV}$ and an exposure of ${\rm TM_T}\gtrsim500$ g yr, we would impose both upper and lower bound at $2\sigma$ or at $3\sigma$ confidence level with a fiducial neutrino lifetime of $\hat{\tau}_{\nu_1}\sim t_0$.

For the more massive lightest neutrinos, the required energy resolution will be relaxed since the $\beta$-decay endpoint is far away from the C$\nu$B signal. 
In Table \ref{tb:2sigmaIH0}, we also show upper limits on the lifetime of $\nu_1$ at $1\sigma$ confidence level for $\nu_{1} \rightarrow \nu_4\phi$ with fiducial values of $\hat{\tau}_{\nu_1}= t_0$ and $\hat{\tau}_{\nu_1}= \infty$, and an exposure of $500$ g yr, considering various values of the lightest neutrino masses $m_{\nu_3}$. The upper limits for any fiducial values of $\hat{\tau}_{\nu_1}\ll t_0$ are almost the same since the results for $\hat{\tau}_{\nu_1}= t_0$ and $\hat{\tau}_{\nu_1}= 0.5t_0$ are almost the same as in Fig.~\ref{fig:chi2_decay0_IH}.
We find that for larger values of $m_{\nu_3}\gtrsim 50\ {\rm meV}$ with an exposure of $500$ g yr, the required energy resolution to obtain constraints on a neutrino lifetime with its fiducial values of $\hat{\tau}_{\nu_1}\lesssim t_0$ and $\hat{\tau}_{\nu_1}= \infty$ at $1\sigma$ confidence level is
\begin{align}
\Delta \lesssim m_{\nu_3}\ \ \  {\rm for}\ \ \  m_{\nu_3}\gtrsim50\ {\rm meV}.
\label{RE1}
\end{align}
This is because the actual $\beta$-decay endpoint energy for the massive lightest neutrino would be shifted by $-m_{\rm lightest}+2\Delta$ from $E_{\rm end}$ for the massless lightest neutrinos in Eq.~(\ref{endpoint}) while the energies for the degenerate C$\nu$B signals are shifted by $E_{\nu_i}\simeq m_{\nu_3}$ from $E_{\rm end}$ for the massless lightest neutrinos (see also Eq.~(\ref{relationEeEnu}) and figures in section~\ref{Sec5.1}).

\begin{table}[h]
\begin{center}
	\begin{tabular}{cccc|cc|cc}
		\hline \hline
	      Ordering & $m_{\nu_3}$ ({\rm meV}) & $\Delta\ ({\rm meV}) $ & ${\rm TM_T}$ (g yr) & $\hat{\tau}_{\nu_1}$  & $1\sigma(\hat{\tau}_{\nu_1}= t_0)$ & $\hat{\tau}_{\nu_1}$ & $1\sigma(\hat{\tau}_{\nu_1}=\infty)$ \\
		\hline 
		 \multirow{4}{*}{IO} &0 & 40 &  \multirow{4}{*}{$500$} & \multirow{4}{*}{$ t_0$} & $\tau_{\nu_1}\lesssim 4t_0$ & \multirow{4}{*}{$\infty$} & $t_0\lesssim \tau_{\nu_1}$  \\
		& 50 & 60 & & & $\tau_{\nu_1}\lesssim 3t_0$ && $t_0\lesssim \tau_{\nu_1}$ \\
		 &100 & 100  & & & $ \tau_{\nu_1}\lesssim 3t_0$ && $t_0\lesssim \tau_{\nu_1}$ \\
		& 200 & 200 & & & $\tau_{\nu_1}\lesssim 3t_0$ && $2t_0\lesssim \tau_{\nu_1}$ \\
		\hline \hline
	\end{tabular}
	\caption{\small{Constraints on the neutrino lifetime $\tau_{\nu_1}$ at 1$\sigma$ confidence level for $\nu_{1} \rightarrow \nu_4\phi$ with fiducial values of $\hat{\tau}_{\nu_1}= t_0$ and $\hat{\tau}_{\nu_1}= \infty$ in the IO case and for various values of the lightest neutrino mass, $m_{\nu_3}$, and of the detector resolution, $\Delta$. We assume an exposure of $500$ g yr. For $m_{\nu_3}\gtrsim 50\ {\rm meV}$, the required energy resolution to obtain constraints at 1$\sigma$ confidence level is $\Delta\lesssim m_{\nu_3}$. }}
  \label{tb:2sigmaIH0}
\end{center}
\end{table}

\clearpage
%%%%%%%%%%%%%%%%%%%%%%%%%%%%%%%%%%%%%%%%%%%%%%%%%%%%%%%%%%%%%%%%%%%%%%%%%%%%%%%%%%%%%%%%%%%%%%%%%%

\subsubsection{2-body decays: $\nu_{3} \rightarrow \nu_1\phi$ and $\nu_{1} \rightarrow \nu_3\phi$}
\label{Sec:5.2.4}

Next we forecast the constraints on the neutrino lifetime and the invisible particle mass in the 2-body decays, $\nu_{3} \rightarrow \nu_1\phi$ for the NO case and $\nu_{1} \rightarrow \nu_3\phi$ for the IO case.

For the IO case, $\nu_{1} \rightarrow \nu_3\phi$, as in Fig.~\ref{fig:Unstable2_Inverse}, the electron spectrum for $\nu_3$ produced by the decays of $\nu_1$ is extremely suppressed, compared to that for $\nu_1$ produced by the decays of $\nu_3$ for the NO case, and the constraint is almost the same with those given in Fig.~\ref{fig:chi2_decay0_IH} for the case with $\nu_1\rightarrow \nu_4\phi$ unless an extremely large exposure is obtained.

\begin{figure}[h]
\begin{minipage}{0.5\hsize}
  \begin{center}
   \includegraphics[clip,width=85mm]{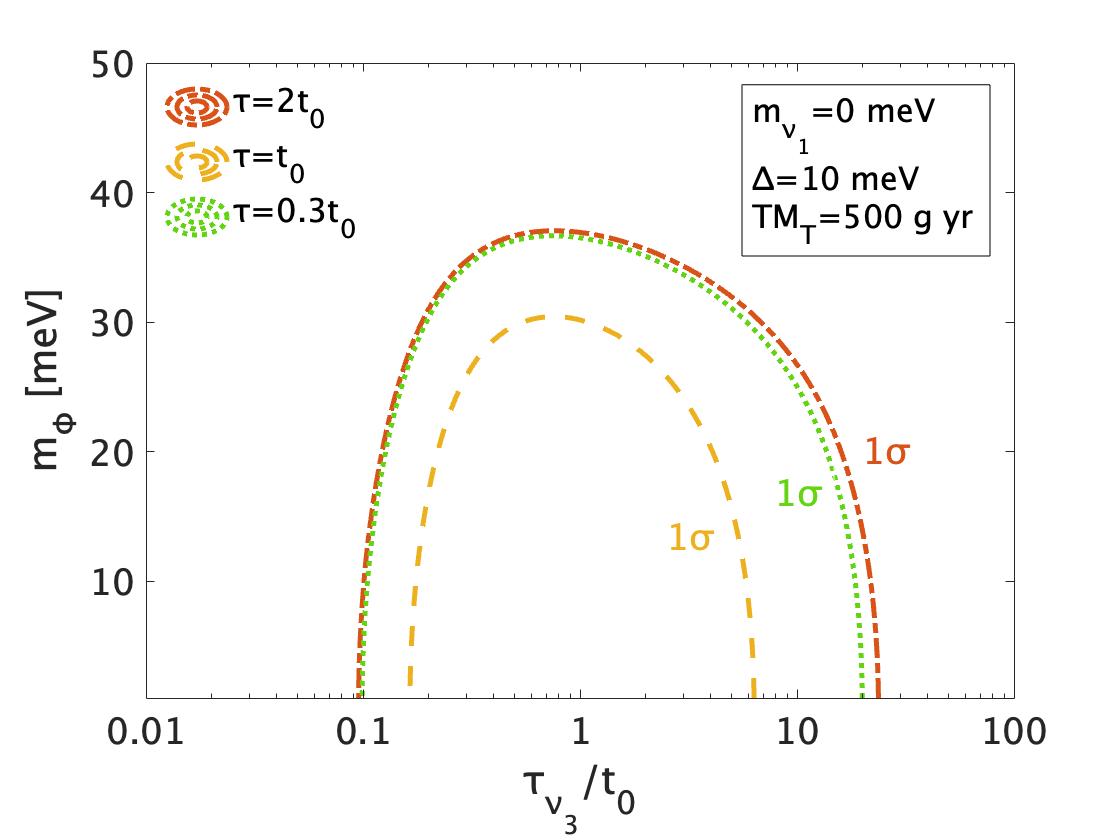}
  \end{center}
\end{minipage}
 \begin{minipage}{0.5\hsize}
  \begin{center}
   \includegraphics[width=85mm]{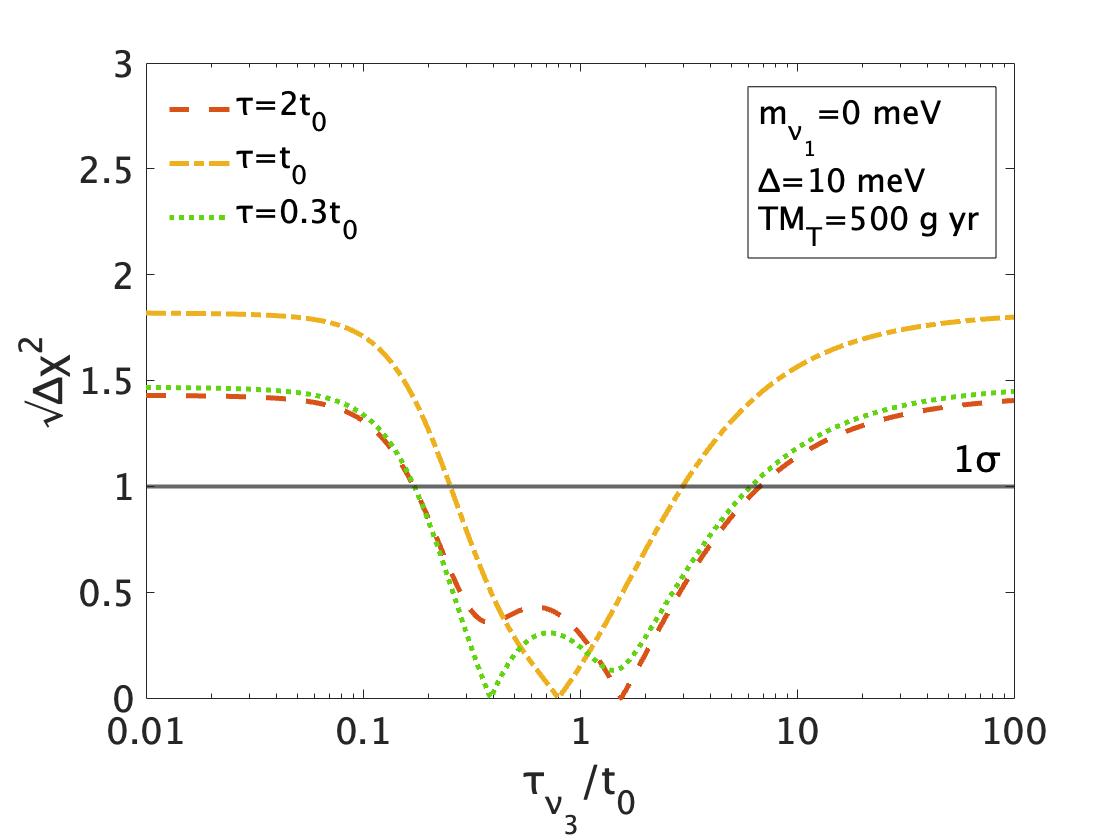}
  \end{center}
  \end{minipage}
  \vspace{-4mm}
 \caption{\small{\underline{Left panels}: Contour plots showing $1\sigma$ confidence intervals in the plane of ($\tau_{\nu_3}$,$m_{\phi}$) for $\nu_{3} \rightarrow \nu_1\phi$ in the NO case with $m_{\nu_1}=0\ {\rm meV}$, $\Delta=10\ {\rm meV}$ and an exposure of ${\rm TM_T}=500$ g yr. 
 \underline{Right panels}: Marginalized $\sqrt{\Delta\chi^2}$, $\sqrt{\Delta \chi^2_{\rm 1-D}}$, as a function of $\tau_{\nu_3}$.
 A fiducial value of ($\hat{\tau}_{\nu_3}$, $\hat{m}_{\phi}$) is assumed to be $(2t_0, 0)$ (red dashed), $(t_0, 0)$ (yellow dot-dashed), and $(0.3t_0, 0)$ (green dotted).}}
 \label{fig:chi2_decay2_NH}
\end{figure}

For the NO case, $\nu_{3} \rightarrow \nu_1\phi$, we show in the left panels of Fig.~\ref{fig:chi2_decay2_NH} the $1\sigma$ confidence intervals in the $(\tau_{\nu_3}, m_{\phi})$ plane for three different fiducial values: $(\hat{\tau}_{\nu_3}, \hat{m}_{\phi})=(2t_0, 0)$ (red dashed), $(t_0, 0)$ (yellow dot-dashed), $(0.3t_0, 0)$ (green dotted). In this case, we assume $m_{\nu_1}=0\ {\rm meV}$, a tiny energy resolution of $\Delta=10\ {\rm meV}$ and an exposure of ${\rm TM_T}=500$ g yr. Due to the large events of the signal for $\nu_1$ from the decays of $\nu_3$ with large $|U_{e1}|$, the constraints on $\tau_{\nu_3}$ become more severe than those for the NO case in the previous section \ref{Sec:5.2.3}. In the case of $m_{\nu_1}=0\ {\rm meV}$, from Fig.~\ref{fig:chi2_decay2_NH}, we find that the upper bound on $m_{\phi}$, $m_{\phi}\lesssim30\ {\rm meV}$, is imposed in addition to the bounds on the lifetime of $\nu_3$, $0.3t_0\lesssim\tau_{\nu_3}\lesssim3t_0$, at $1\sigma$ confidence level for a fiducial value of $(\hat{\tau}_{\nu_3}, \hat{m}_{\phi})=(t_0,0)$. 
The constraints on $(\hat{\tau}_{\nu_3}, \hat{m}_{\phi})=(2t_0, 0)$ and $(0.3t_0, 0)$ are almost the same since the less signal for $\nu_1$ from the decays of $\nu_3$ is produced with longer $\tau_{\nu_3}$ while the energy of $\nu_1$ from the decays of $\nu_3$ with the shorter lifetime is diluted by the longer cosmic expansion.
In other words, in the two cases of $(\hat{\tau}_{\nu_3}, \hat{m}_{\phi})=(2t_0, 0)$ and $(0.3t_0, 0)$, the distinguishable number of events for $\nu_1$ produced by the decays of $\nu_3$ from the background are almost the same.
In the right panel of Fig.~\ref{fig:chi2_decay2_NH}, the marginalized $\sqrt{\Delta\chi^2}$ as a function of $\tau_{\nu_3}$ is shown.
In both the cases of $(\hat{\tau}_{\nu_3}, \hat{m}_{\phi})=(2t_0, 0)$ and $(0.3t_0, 0)$, $\sqrt{\Delta \chi^2}$ contains a local minimum in the opposite true minimum each other for the same reason as above.
For $(\hat{\tau}_{\nu_3}, \hat{m}_{\phi})=(t_0, 0)$, $\sqrt{\Delta \chi^2}$ contains no local minimum since the distinguishable number of events for $\nu_1$ produced by the decays of $\nu_3$ from the background would be maximum in the case of $(m_{\nu_1}, \Delta, {\rm TM_T})=(0, 10\ {\rm meV}, 500\ {\rm g\ yr})$.

\begin{figure}[h]
  \begin{center}
   \includegraphics[clip,width=85mm]{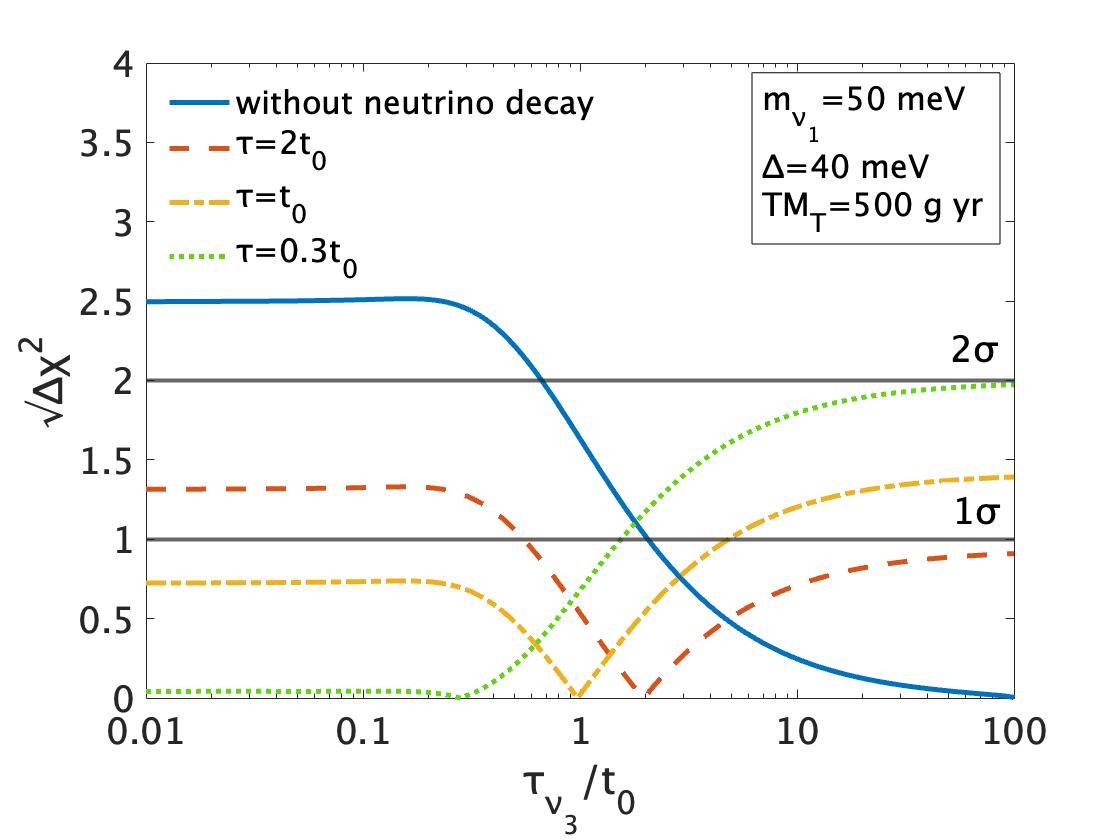}
  \end{center}
  \vspace{-6mm}
 \caption{\small{Marginalized $\sqrt{\Delta\chi^2}$, $\sqrt{\Delta \chi^2_{\rm 1-D}}$, as a function of $\tau_{\nu_3}$ for $\nu_{3} \rightarrow \nu_1\phi$ in the NO case with $m_{\nu_1}=50\ {\rm meV}$, $\Delta=40\ {\rm meV}$ and an exposure of ${\rm TM_T}=500$ g yr.
 The fiducial value of ($\hat{\tau}_{\nu_3}$, $\hat{m}_{\phi}$) is assumed to be $(\infty, 0)$ (blue solid),  $(2t_0, 0)$ (red dashed), $(t_0, 0)$ (yellow dot-dashed), and $(0.3t_0, 0)$ (green dotted).
 Marginalized $\sqrt{\Delta\chi^2}$ as a function of $\tau_{\nu_3}$ for $\nu_{3} \rightarrow \nu_1\nu_4\bar{\nu}_4$ in the NO case with $m_{\nu_1}=50\ {\rm meV}$, $\Delta=40\ {\rm meV}$ and an exposure of ${\rm TM_T}=500$ g yr are almost the same with those given in this figure.
 }}
 \label{fig:chi2_decay02_NH_nu50}
\end{figure}

In Fig.~\ref{fig:chi2_decay02_NH_nu50}, we show the marginalized constraints on $\tau_{\nu_3}$ for $m_{\nu_1}=50\ {\rm meV}$. In this case, constraints on $m_{\phi}$ are not obtained. Note that in the case of $m_{\nu_1}=50\ {\rm meV}$, $m_{\phi}\lesssim20\ {\rm meV}$ is allowed by the energy conservation. For $m_{\nu_1}=50\ {\rm meV}$, even in an energy resolution of $\Delta=40\ {\rm meV}$, we see the bounds on a neutrino lifetime at $1\sigma$, $\tau_{\nu_3}\lesssim2t_0$, for the fiducial value of $(\hat{\tau}_{\nu_3}, \hat{m}_{\phi})=(0.3t_0,0)$

In Table \ref{tb:1sigmaNH2}, we also show marginalized constraints on the neutrino lifetime $\tau_{\nu_3}$ at $1\sigma$ confidence level for $\nu_{3} \rightarrow \nu_1\phi$ with fiducial values of $\hat{\tau}_{\nu_3}=t_0$ and $\hat{\tau}_{\nu_3}=\infty$, and an exposure of $500$ g yr, considering various values of the lightest neutrino masses $m_{\nu_1}$. For $\hat{\tau}_{\nu_3}=\infty$, we cannot find a lower bound on $\tau_{\nu_3}$ since for $\tau_{\nu_3}\ll t_0$ the signal for $\nu_1$ produced by the decay of $\nu_3$ is too diluted by cosmic expansion to distinguish from the $\beta$-decay background. For a fiducial value of $\hat{\tau}_{\nu_3}\ll t_0$, the upper limits with $m_{\nu_1}\gtrsim 50\ {\rm meV}$ do not change drastically since the electron spectrum for $\nu_1$ produced by the decays of $\nu_3$ is Gaussian centered at $E_{e}\simeq K_{\rm end}^0+m_e+m_{\nu_1}$ (see the bottom panels in Fig.~\ref{fig:Unstable2_phi}). On the other hand, the constraint on $\tau_{\nu_3}$ for a fiducial value of $\hat{\tau}_{\nu_3}\ll t_0$ with $m_{\nu_1}\simeq 0\ {\rm meV}$ become weaker because the energy injection into $\nu_1$ from the decays of $\nu_3$ is enough diluted and the spectrum produced by the decays of $\nu_3$ is at $E_{e}\sim K_{\rm end}^0+m_e$. The required energy resolution to obtain constraints on $\tau_{\nu_3}$ with its fiducial value of $\hat{\tau}_{\nu_3}\lesssim t_0$ and $\hat{\tau}_{\nu_3}=\infty$ at $1\sigma$ confidence level with $m_{\nu_1}\gtrsim 50\ {\rm meV}$ and an exposure of 500 g yr is
 \begin{align}
 \Delta \lesssim 0.6m_{\nu_1}\ \ \ {\rm for}\ \ \  m_{\nu_1}\gtrsim50\ {\rm meV}.
 \end{align}
This resolution is smaller than that given in Eq.~(\ref{RE1}) since the signal for massive $\nu_1$ produced by the decays of $\nu_3$ is overlapped with that for massive $\nu_1$ produced in the early universe in the NO case. 
%Here we neglect the gravitational clustering in the spectrum for $\nu_1$ produced by the decays of $\nu_3$. If we take this effect into account, we may impose more severe constraints on a neutrino lifetime and $m_{\phi}$.
 
\begin{table}[h]
\begin{center}
	\begin{tabular}{cccc|cc|cc}
		\hline \hline
	      Ordering & $m_{\nu_1}$ ({\rm meV}) & $\Delta\ ({\rm meV}) $ & ${\rm TM_T}$ (g yr) & $\hat{\tau}_{\nu_3}$  & $1\sigma(\hat{\tau}_{\nu_3}=t_0)$ & $\hat{\tau}_{\nu_3}$  & $1\sigma(\hat{\tau}_{\nu_3}=\infty)$ \\
		\hline 
		 \multirow{4}{*}{NO} &0 & 10 &  \multirow{4}{*}{$500$} & \multirow{4}{*}{$t_0$} & $0.2t_0\lesssim\tau_{\nu_3}\lesssim 3t_0$ & \multirow{4}{*}{$\infty$} & \small{No bound} \\
		& 50 & 40 & &  & $\tau_{\nu_3}\lesssim 5t_0$ && $2t_0\lesssim \tau_{\nu_3} $ \\
		 &100 & 60  & & & $\tau_{\nu_3}\lesssim 3t_0$ && $4t_0\lesssim \tau_{\nu_3} $\\
		 &200 & 120  & & & $\tau_{\nu_3}\lesssim 7t_0$ && $6t_0\lesssim \tau_{\nu_3} $ \\
		\hline \hline
	\end{tabular}
	\caption{\small{Marginalized constraints on the neutrino lifetime $\tau_{\nu_3}$ at 1$\sigma$ confidence level for $\nu_{3} \rightarrow \nu_1\phi$ with fiducial values of $\hat{\tau}_{\nu_3}=t_0$ and $\hat{\tau}_{\nu_3}=\infty$ in the NO case and for various values of the lightest neutrino mass, $m_{\nu_1}$, and of the detector resolution, $\Delta$. We assume an exposure of $500$ g yr.
 }}
  \label{tb:1sigmaNH2}
\end{center}
\end{table}

\clearpage
%%%%%%%%%%%%%%%%%%%%%%%%%%%%%%%%%%%%%%%%%%%%%%%%%%%%%%%%%%%%%%%%%%%%%%%%%%%%%%%%%%%%%%%%%%%%%%%%%%

\subsubsection{3-body decays: $\nu_3\rightarrow \nu_1\nu_4\bar{\nu}_4$ and $\nu_1 \rightarrow \nu_3\nu_4\bar{\nu}_4$}
\label{Sec:5.2.5}

Finally we forecast the constraints on the neutrino lifetime and the invisible particle mass in the 3-body decays. For simplicity, we only consider the following 3-body decays,  $\nu_3 \rightarrow \nu_1\nu_4\bar{\nu}_4$ for the NO case. For $\nu_1 \rightarrow \nu_3\nu_4\bar{\nu}_4$ in the IO case, due to the suppressed spectrum of $\nu_3$ from the decays of $\nu_1$ as in Fig.~\ref{fig:Unstable3_Inverse}, the result is also the same with Fig.~\ref{fig:chi2_decay0_IH} unless a very high exposure is obtained.

\begin{figure}[h]
\begin{minipage}{0.5\hsize}
  \begin{center}
   \includegraphics[clip,width=85mm]{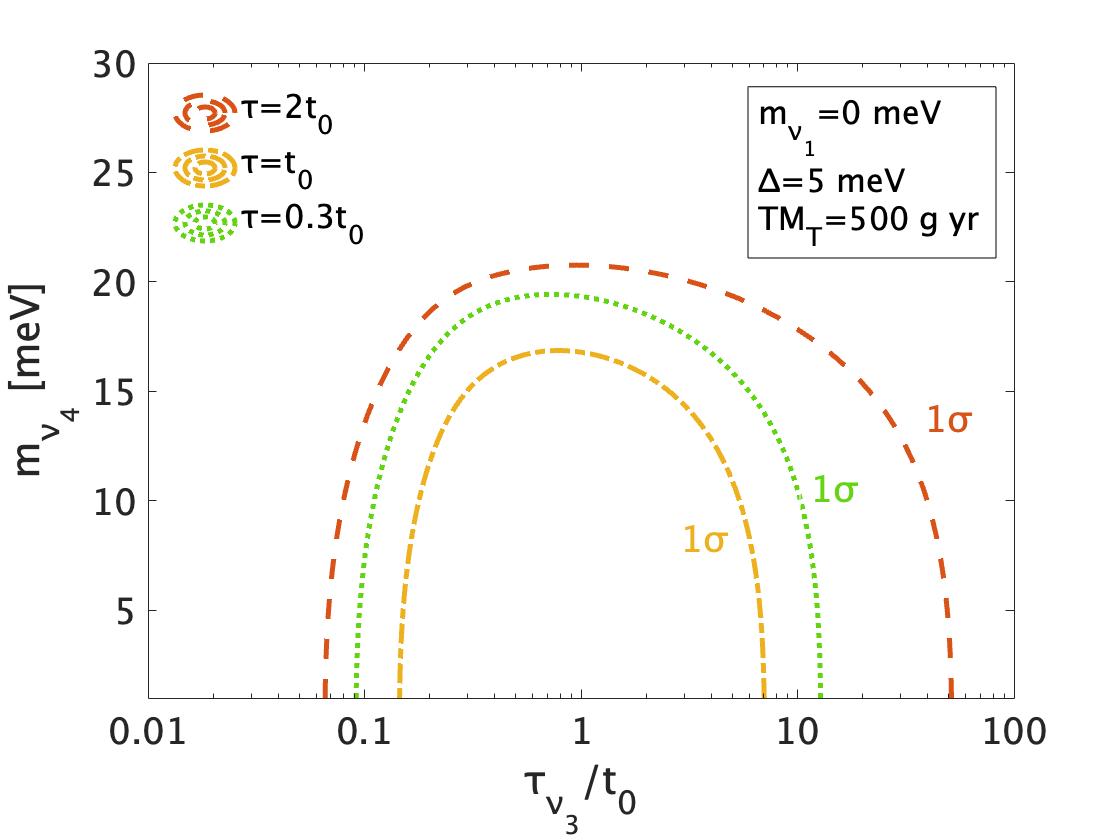}
  \end{center}
\end{minipage}
 \begin{minipage}{0.5\hsize}
  \begin{center}
   \includegraphics[width=85mm]{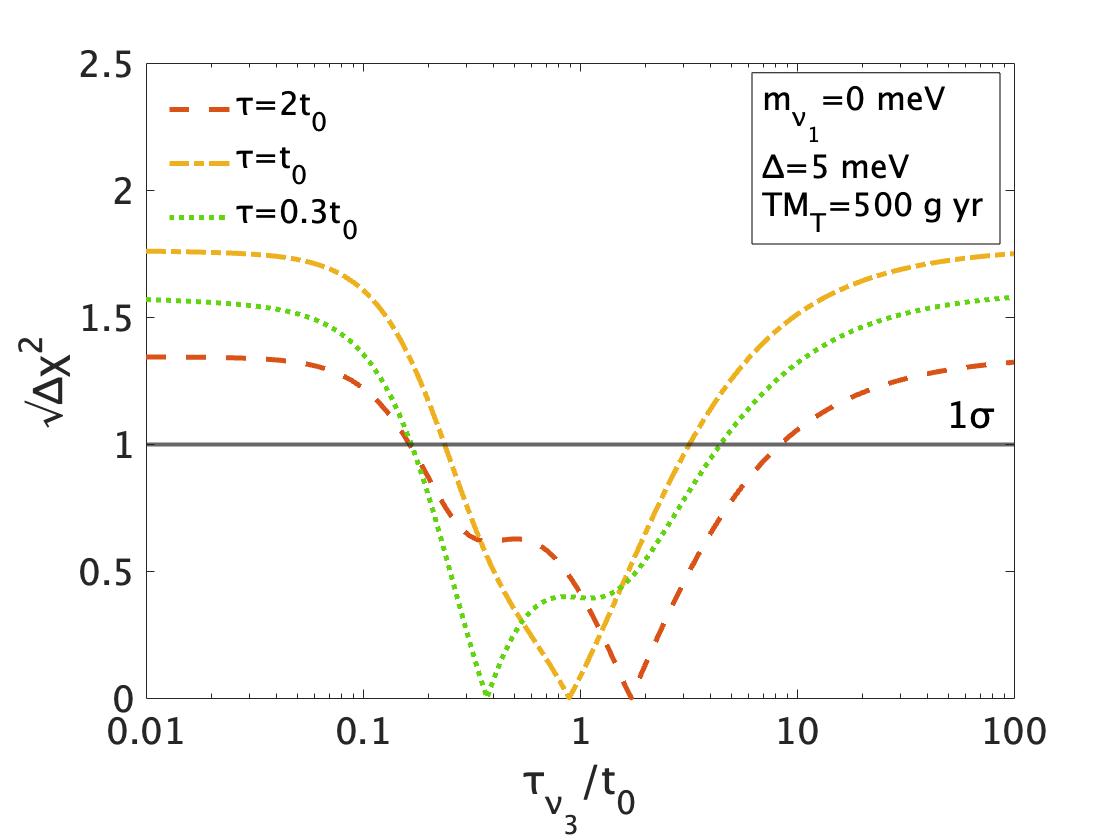}
  \end{center}
  \end{minipage}
  \vspace{-4mm}
 \caption{\small{\underline{Left panels}: Contour plots showing $1\sigma$ confidence intervals in the plane of ($\tau_{\nu_3}$,$m_{\phi}$) for $\nu_{3} \rightarrow \nu_1\nu_4\bar{\nu}_4$ in the NO case with $m_{\nu_1}=0\ {\rm meV}$ and $\Delta=5\ {\rm meV}$. 
 \underline{Right panels}: Marginalized $\sqrt{\Delta\chi^2}$,$\sqrt{\Delta \chi^2_{\rm 1-D}}$, as a function of $\tau_{\nu_3}$.
 A fiducial value of ($\hat{\tau}_{\nu_3}$, $\hat{m}_{\phi}$) is assumed to be $(2t_0, 0)$ (red dashed), $(t_0, 0)$ (yellow dot-dashed), and $(0.3t_0, 0)$ (green dotted).
 An exposure of ${\rm TM_T}=500$ g yr is also assumed.}}
 \label{fig:chi2_decay3_NH}
\end{figure}

For the NO case, $\nu_3 \rightarrow \nu_1\nu_4\bar{\nu}_4$, we show in Fig.~\ref{fig:chi2_decay3_NH}  the $1\sigma$ confidence intervals in the $(\tau_{\nu_3}, m_{\nu_4})$ plane and the marginalized constraints on $\tau_{\nu_3}$ for three different fiducial values: $(\hat{\tau}_{\nu_3}, \hat{m}_{\nu_4})=(2t_0, 0)$ (red dashed), $(t_0, 0)$ (yellow dot-dashed), $(0.3t_0, 0)$ (green dotted).
In this case, we assume $m_{\nu_1}=0\ {\rm meV}$, a high energy resolution of $\Delta=5\ {\rm meV}$ and an exposure of ${\rm TM_T}=500$. 
Since the spectrum for $\nu_1$ produced by the decays of $\nu_3$ decreases around the maximum energy of emitted $\nu_1$ in the 3-body decays, $E_{\nu_1}^{\rm max}$ in Eq.~(\ref{Emax3}), the constraints on $\tau_{\nu_3}$ for $m_{\nu_1}\simeq 0\ {\rm meV}$ become weaker and the smaller energy resolution is required compared with the 2-body decays.
For a fiducial value of $(\hat{\tau}_{\nu_3}, \hat{m}_{\nu_4})=(t_0, 0)$ in the case of $m_{\nu_1}= 0\ {\rm meV}$, we find that the upper bounds on $m_{\nu_4}$, $m_{\nu_4}\lesssim17\ {\rm meV}$, are imposed while the bounds on $\tau_{\nu_3}$ are $0.1t_0\lesssim\tau_{\nu_3}\lesssim7t_0$. We also find that the marginalized constraints on $\tau_{\nu_3}$ for $m_{\nu_1} = 50\ {\rm meV},\ \Delta=40\ {\rm meV}$ and ${\rm TM_T}=500$ g yr  are almost the same with those given in Fig.~\ref{fig:chi2_decay02_NH_nu50}. In this case, the constraint on $m_{\nu_4}$ is not obtained. Note that in the case of $m_{\nu_1}=50\ {\rm meV}$, $m_{\nu_4}\lesssim10\ {\rm meV}$ is allowed by the energy conservation.

In Table~\ref{tb:1sigmaNH3}, we also show marginalized constraints on the neutrino lifetime $\tau_{\nu_3}$ at $1\sigma$ confidence level for $\nu_{3} \rightarrow \nu_1\nu_4\bar{\nu}_4$ with fiducial values of $\hat{\tau}_{\nu_3}=t_0$ and $\hat{\tau}_{\nu_3}=\infty$, and an exposure of $500$ g yr, considering various values of the lightest neutrino masses $m_{\nu_1}$.
For $m_{\nu_1}\gtrsim 50\ {\rm meV}$, the results are almost the same with the case of $\tau_{\nu_3}$ for $\nu_{3} \rightarrow \nu_1\phi$ as in Table \ref{tb:1sigmaNH2} since the electron spectrum for $\nu_1$ produced by the decays of $\nu_3$ is Gaussian centered at $E_{e}\simeq K_{\rm end}^0+m_e+m_{\nu_1}$ and does not depend on the decay process approximately.

To distinguish 2-body decays and 3-body decays, more exposure and a better energy resolution may be needed since in general, there are many undetermined parameters in both cases. We leave this analysis to future work.

\begin{table}[h]
\begin{center}
	\begin{tabular}{cccc|cc|cc}
		\hline \hline
	      Ordering & $m_{\nu_1}$ ({\rm meV}) & $\Delta\ ({\rm meV}) $ & ${\rm TM_T}$ (g yr) & $\hat{\tau}_{\nu_3}$  & $1\sigma(\hat{\tau}_{\nu_3}=t_0)$ & $\hat{\tau}_{\nu_3}$  & $1\sigma(\hat{\tau}_{\nu_3}=\infty)$ \\
		\hline 
		 \multirow{4}{*}{NO} &0 & 5 &  \multirow{4}{*}{$500$} & \multirow{4}{*}{$t_0$} & $0.2t_0\lesssim\tau_{\nu_3}\lesssim 3t_0$ & \multirow{4}{*}{$\infty$} & \small{No bound} \\
		& 50 & 40 & &  & $\tau_{\nu_3}\lesssim 5t_0$ && $2t_0\lesssim \tau_{\nu_3}$  \\
		 &100 & 60  & & & $\tau_{\nu_3}\lesssim 3t_0$ && $4t_0\lesssim \tau_{\nu_3}$ \\
		 &200 & 120  & & & $\tau_{\nu_3}\lesssim 7t_0$ && $6t_0\lesssim \tau_{\nu_3}$ \\
		\hline \hline
	\end{tabular}
	\caption{\small{Marginalized constraints on the neutrino lifetime $\tau_{\nu_3}$ at 1$\sigma$ confidence level for $\nu_{3} \rightarrow \nu_1\nu_4\bar{\nu}_4$ with fiducial values of $\hat{\tau}_{\nu_3}=t_0$ and $\hat{\tau}_{\nu_3}=\infty$ in the NO case and for various values of the lightest neutrino mass, $m_{\nu_1}$, and of the detector resolution, $\Delta$.
	We assume an exposure of $500$ g yr.
	 }}
  \label{tb:1sigmaNH3}
\end{center}
\end{table}

\clearpage
%%%%%%%%%%%%%%%%%%%%%%%%%%%%%%%%%%%%%%%%%%%%%%%%%%%%%%%%%%%%%%%%%%%%%%%%%%%%%%%%%%%%%%%%%%%%%%%%%%
\subsubsection{Relation between confidence level and exposure of tritium}
\label{sec:5.2.5}

In previous sections we mainly forecast constraints on neutrino lifetimes at $1\sigma$ confidence level in cosmic neutrino capture experiment on tritium. In this section we shortly discuss on relation between constraints on neutrino lifetime at $2\sigma$ and more confidence level and the required exposure of tritium. 

In our definition of $\chi^2(\bm{\theta})$ given in Eq.~(\ref{chi2}), $\chi^2$ is proportional to the exposure time $\mathrm{T}$ and the total mass of tritium $\mathrm{M_T}$,
\begin{align}
    \chi^2(\bm{\theta}) \propto \mathrm{TM_T}.
\end{align}
The $\chi^2$ marginalized over $m_{\phi, \nu_4}$ as defined in Eq.~(\ref{Mchi2}) is estimated approximately
\begin{align}
    \mathcal{L}_{\rm 1-D}&=\int dm \exp\left[-\frac{1}{2}\chi^2(\bm{\theta}) \right] \nonumber \\
    &\sim dm \exp\left[-\frac{1}{2}\underset{m}{\min}\chi^2(\bm{\theta}) \right], \nonumber \\
    \chi_{\rm 1-D}^2(\tau_{\nu_i})&=-2\log \mathcal{L}_{\rm 1-D}, \nonumber \\
    &\sim \underset{m}{\min}\chi^2(\bm{\theta}),
\end{align}
where we neglect $dm$ since $dm$ is canceled out in $\Delta \chi_{\rm 1-D}^2$.
Then $\chi^2_{\rm 1-D}(\tau_{\nu_i})$ is also proportional to ${\rm TM_T}$.

In the previous sections, we mainly assume $500$ g yr of tritium with a fixed $\Delta$ and then constrain neutrino lifetime at $1\sigma$ or slightly higher confidence level. 
When we use $\mathrm{TM_T}$ of tritium, the confidence level for marginalized constraints on neutrino lifetimes is expected to be enhanced by $(\mathrm{TM_T}/500\ {\rm g\ yr})^{1/2}$ compared with $500\ {\rm g\ yr}$ of tritium.
If we use $2000$ g yr of tritium and the same energy resolutions in the previous sections, it is expected to impose marginalized constraints on neutrino lifetimes at $2\sigma$ or $3\sigma$ confidence level in some cases.
We have numerically confirmed that we obtain similar constraints on neutrino lifetimes at $2\sigma$ confidence level with 2000 g yr of tritium if we obtain the constraints at $1\sigma$ confidence level with 500 g yr of tritium in all cases of section \ref{Sec5.2}~\footnote{Using the Poissonian likelihood (\ref{Plikelihood}), we have also confirmed that similar constraints at $2\sigma$ confidence level are obtained.}.

%%%%%%%%%%%%%%%%%%%%%%%%%%%%%%%%%%%%%%%%%%%%%%%%%%%%%%%%%%%%%%%%%%%%%%%%%%%%%%%%%%%%%%%%%%%%%%%%%%
%%%%%%%%%%%%%%%%%%%%%%%%%%%%%%%%%%%%%%%%%%%%%%%%%%%%%%%%%%%%%%%%%%%%%%%%%%%%%%%%%%%%%%%%%%%%%%%%%%
%%%%%%%%%%%%%%%%%%%%%%%%%%%%%%%%%%%%%%%%%%%%%%%%%%%%%%%%%%%%%%%%%%%%%%%%%%%%%%%%%%%%%%%%%%%%%%%%%%
%%%%%%%%%%%%%%%%%%%%%%%%%%%%%%%%%%%%%%%%%%%%%%%%%%%%%%%%%%%%%%%%%%%%%%%%%%%%%%%%%%%%%%%%%%%%%%%%%%
\section{Conclusions}
\label{Sec6}

Future direct observations of the C$\nu$B will be able to impose significant constraints on a neutrino lifetime, especially in the region of the age of the universe, $t_0=4.35\times 10^{17}\ {\rm s}$, once we would obtain enough statistics on the signal of the C$\nu$B. In sections~\ref{Sec3} and \ref{Sec4}, we have formulated invisible neutrino decay rates and lighter neutrino spectra from heavier neutrino decays for the 2-body and 3-body decays in the current universe, respectively. After giving the formulation, in section~\ref{Sec5}, we have forecasted the would-be observed spectra for unstable neutrinos, and the constraints on a neutrino lifetime and an invisible particle mass in cosmic neutrino capture on tritium, in particular, the PTOLEMY-type experiment.
The main forecasts of the constraints on neutrino decays are summarized as follows:
\begin{itemize}
  \setlength{\parskip}{0mm}
  \setlength{\itemsep}{1mm}
  \item For any processes of decays of $\nu_3$ in the NO case and $\nu_1$ in the IO case, constraints on a neutrino lifetime will be imposed as in section \ref{Sec:5.2.3}. 
In the NO case, the neutrino lifetime $\tau_{\nu_3}$ for the case without neutrino decays is bounded to be $\tau_{\nu_3}\gtrsim0.1t_0$ at $1\sigma$ confidence level with a large exposure of 1000 g yr and the detector resolution of $20\ {\rm meV}$ (see Fig.~\ref{fig:chi2_decay0_NH}). 
In the IO case, for $m_{\nu_3}\gtrsim50 \rm meV\ (m_{\nu_3}\lesssim 50\ {\rm meV})$, the neutrino lifetime $\tau_{\nu_1}$ with fiducial values of $\hat{\tau}_{\nu_1}\lesssim t_0$ and $\hat{\tau}_{\nu_1}=\infty$ will be bounded to be $\tau_{\nu_1}\lesssim 4t_0$ and $t_0\lesssim\tau_{\nu_1}$, respectively, at $1\sigma$ confidence level with an exposure of ${\rm TM_T}\gtrsim 500$ g yr and the detector resolution of $\Delta\lesssim m_{\nu_3} \ (\Delta \lesssim 40\ {\rm meV})$ (see Table~\ref{tb:2sigmaIH0}). 
  
  \item For the 2-body decay process in the NO case, $\nu_{3} \rightarrow \nu_1\phi$, as in section \ref{Sec:5.2.4}, the neutrino lifetime $\tau_{\nu_3}$ with a fiducial value of $\hat{\tau}_{\nu_3}\simeq t_0$ will be constrained to be $0.2t_0\lesssim \tau_{\nu_3} \lesssim 3t_0$ at $1\sigma$ confidence level with $m_{\nu_1}=0\ {\rm meV}$, an exposure of $500$ g yr and the resolution of $10\ {\rm meV}$ (see Fig.~\ref{fig:chi2_decay2_NH}). In addition, in this case, the constraint on the invisible particle mass, $m_{\phi}\lesssim 40 {\rm meV}$, will also be imposed. 
  For $m_{\nu_1}\gtrsim 50\ {\rm meV}$, even in the resolution of $\Delta \lesssim 0.6m_{\nu_1}\ {\rm meV}$, the constraints on neutrino lifetimes at $1\sigma$ confidence level, $\tau_{\nu_3}\lesssim 10t_0$ and $2t_0 \lesssim \tau_{\nu_3}$ for fiducial values of $\hat{\tau}_{\nu_3}\lesssim t_0$ and $\hat{\tau}_{\nu_3}=\infty$ respectively, will be obtained with an exposure of ${\rm TM_T}\gtrsim 500$ g yr  (see Table~\ref{tb:1sigmaNH2}).
  
  \item For the 3-body decay process in the NO case, $\nu_{3} \rightarrow \nu_1\nu_4\bar{\nu}_4$, as in section \ref{Sec:5.2.5}, the neutrino lifetime $\tau_{\nu_3}$ with a fiducial value $\hat{\tau}_{\nu_3} \simeq t_0$ is bounded to be $0.2t_0\lesssim \tau_{\nu_3} \lesssim 3t_0$ at $1\sigma$ confidence level with an exposure of $500$ g yr and the smaller resolution of $5\ {\rm meV}$. In this case, the constraints on the invisible particle mass, $m_{\nu_4}\lesssim 17\ {\rm meV}$ will be also imposed (see Fig.~\ref{fig:chi2_decay3_NH}). For $m_{\nu_1}\gtrsim 50\ {\rm meV}$, the constraints on the neutrino lifetime $\tau_{\nu_3}$ will be almost the same with the 2-body decays for $\nu_{3} \rightarrow \nu_1\phi$ (see Table~\ref{tb:1sigmaNH3}, Fig.~\ref{fig:Unstable2_phi} and \ref{fig:Unstable3_nu0}).
  
\end{itemize}
In the case of the decays in the NO case, $\nu_{3} \rightarrow \nu_1$, the total event number of the C$\nu$B is larger than the case without neutrino decays thanks to large $|U_{e1}|$. These neutrino decays may make it easier for us to detect the C$\nu$B in the PTOLEMY-type experiment.
Here we assume $\nu_2$ are stable and neglect gravitational clustering by our Galaxy and nearby galaxies in lighter neutrinos produced by the decays. If $\nu_2$ also decays, the constraints on $\tau_{\nu_3}$ might be better since the background fluctuation from $\nu_2$ is reduced. In addition, the gravitational attraction by our Galaxy and nearby galaxies in neutrinos produced by the decays might enhance statistics on neutrino decays.

The required energy resolutions to obtain the constraints on the neutrino lifetime at $1\sigma$ confidence level are almost the same as the required resolution to discover the C$\nu$B signal at the confidence level of $1\sigma$ or higher in the case without neutrino decays (see Fig.~5 in ref.~\cite{Betti:2019ouf}), but we neglect a tiny constant background noise in this work.
An exposure of 500 g yr can be achieved in the PTOLEMY-type experiment with 100 g of tritium after 5-year data taking.
To improve the confidence level for neutrino decays, more exposure will be necessary as discussed in section \ref{sec:5.2.5}. 
A more quantitative discussion will be possible after the more concrete PTOLEMY setup is decided, and the neutrino mass ordering and the lightest neutrino mass are constrained more severely from complementary future neutrino experiments.

To summarize, the PTOLEMY-type experiment would be able to constrain neutrino decays with long lifetimes especially for the cases that the lightest neutrino mass is not too small or the neutrino mass ordering is inverted if enough small energy resolution and large exposure are obtained.  
On the other hand, for the normal ordering case with $m_{\rm lightest}=0$, it would be quite difficult to impose constraints on neutrino decays for any channels unless one would realize a very small energy resolution and a very large exposure.
A future C$\nu$B experiment may reveal non-standard neutrino interactions with very light particles.

%%%%%%%%%%%%%%%%%%%%%%%%%%%%%%%%%%%%%%%%%%%%%%%%%%%%%%%%%%%%%%%%%%%%%%%%%%%%%%%%%%%%%%%%%%%%%%%%%%
%%%%%%%%%%%%%%%%%%%%%%%%%%%%%%%%%%%%%%%%%%%%%%%%%%%%%%%%%%%%%%%%%%%%%%%%%%%%%%%%%%%%%%%%%%%%%%%%%%
%%%%%%%%%%%%%%%%%%%%%%%%%%%%%%%%%%%%%%%%%%%%%%%%%%%%%%%%%%%%%%%%%%%%%%%%%%%%%%%%%%%%%%%%%%%%%%%%%%
%%%%%%%%%%%%%%%%%%%%%%%%%%%%%%%%%%%%%%%%%%%%%%%%%%%%%%%%%%%%%%%%%%%%%%%%%%%%%%%%%%%%%%%%%%%%%%%%%%

\section*{Acknowledgments}
We are greatful to Miguel Escudero, Jacobo Lopez-Pavon, Nuria Rius and Stefan Sandner for checking a part of the calculations for the decay rates,
Tomo Takahashi and Peter B. Denton for valuable comments on the statistics, and Michiru Niibo for useful comments.
KA is supported by JSPS Grant-in-Aid for Research Fellows
No. 19J14449 and the IBS under project code IBS-R018-D1. GL is supported by INFN and by MIUR. MY is supported in part by JSPS Grant-in-Aid for Scientific Research Numbers JP18K18764, JP21H01080, JP21H00069.

%%%%%%%%%%%%%%%%%%%%%%%%%%%%%%%%%%%%%%%%%%%%%%%%%%%%%%%%%%%%%%%%%%%%%%%%%%%%%%%%%%%%%%%%%%%%%%%%%%
%%%%%%%%%%%%%%%%%%%%%%%%%%%%%%%%%%%%%%%%%%%%%%%%%%%%%%%%%%%%%%%%%%%%%%%%%%%%%%%%%%%%%%%%%%%%%%%%%%
%%%%%%%%%%%%%%%%%%%%%%%%%%%%%%%%%%%%%%%%%%%%%%%%%%%%%%%%%%%%%%%%%%%%%%%%%%%%%%%%%%%%%%%%%%%%%%%%%%
%%%%%%%%%%%%%%%%%%%%%%%%%%%%%%%%%%%%%%%%%%%%%%%%%%%%%%%%%%%%%%%%%%%%%%%%%%%%%%%%%%%%%%%%%%%%%%%%%%

\appendix

\section{Invisible neutrino decay rates}
\label{appa}

In this appendix, we show the formula for the invisible neutrino decay rates, $\Gamma_{\nu_i}$, in the several cases of  2-body and 3-body decays, all of which match with those in appendix A.1 of the latest version of ref.~\cite{Escudero:2020ped}.
In the case of 3-body decays, we also show the decay rates in each neutrino energy produced by the decays, $d \Gamma_{\nu_i\rightarrow \nu_j\nu_k\bar{\nu}_l}/dE_{\nu_j}$, to obtain cosmic neutrino spectra from the decays.
We can apply the formula to both cases where cosmic neutrinos decay into active neutrinos and sterile neutrinos.

\subsection{2-body decays}
\label{appa1}

\subsubsection{Decays into bosons with spin 0: $\nu_i\rightarrow \nu_j\phi$}

In this case, the general interaction Lagrangian between neutrinos and bosons with spin 0, $\phi$, at renormalizable level is given by
\begin{align}
\mathcal{L}^{\phi}_{\rm int}=\phi \bar{\nu}_i(h_{ij}+i\lambda_{ij}\gamma_5)\nu_j+{\rm h.c.},
\label{generalphiinteraction}
\end{align}
where $i,j=1,2,3,4$ and $\nu_4$ denotes sterile neutrinos (or unknown fermions). 

The decay rates for $\nu_i \rightarrow \nu_j\phi$ is given by
\begin{align}
\Gamma_{\nu_i\rightarrow \nu_j\phi}&=\frac{1}{4\pi m_{\nu_i}}\sqrt{\left[m_{\nu_i}^2-(m_{\nu_j}+m_{\phi})^2\right]\left[m_{\nu_i}^2-(m_{\nu_j}-m_{\phi})^2\right]} \nonumber \\
&\ \ \ \  \times \left\{|h_{ij}|^2\left[\left(1+\frac{m_{\nu_j}}{m_{\nu_i}} \right)^2-\frac{m_{\phi}^2}{m_{\nu_i}^2} \right]+|\lambda_{ij}|^2 \left[\left(1-\frac{m_{\nu_j}}{m_{\nu_i}} \right)^2-\frac{m_{\phi}^2}{m_{\nu_i}^2} \right]\right\}.
\end{align}

%%%%%%%%%%%%%%%%%%%%%%%%%%%%%%%%%%%%%%%%%%%%%%%%%%%%%%%%%%%%%%%%%%%%%%%%%%%%%%%%%%%%%%%%%%%%%%%%%%
%%%%%%%%%%%%%%%%%%%%%%%%%%%%%%%%%%%%%%%%%%%%%%%%%%%%%%%%%%%%%%%%%%%%%%%%%%%%%%%%%%%%%%%%%%%%%%%%%%

\subsubsection{Decays into bosons with spin 1: $\nu_i\rightarrow \nu_jZ'$}

The general interaction Lagrangian between neutrinos and bosons with spin 1, $Z'$, at renormalizable level is given by
\begin{align}
\mathcal{L}^{Z'}_{\rm int}=Z'_{\mu}\bar{\nu}_i\gamma^{\mu}(g_{ij}^LP_L+g_{ij}^RP_R)\nu_j+{\rm h.c.},
\label{generalZinteraction}
\end{align}
where $P_{L(R)}=(1-(+)\gamma_5)/2$ are left-(right-) handed projection operator and $g_{ij}^R$ is assumed to be the non-zero only for $g_{44}^R$ (that is, the other components vanish).

Then the decay rates for $\nu_i\rightarrow \nu_jZ'$ is given by 
\begin{align}
\Gamma_{\nu_i\rightarrow \nu_jZ'}&=\frac{|g_{ij}^L|^2}{8\pi m_{\nu_i}}\sqrt{\left[m_{\nu_i}^2-(m_{\nu_j}+m_{Z'})^2\right]\left[m_{\nu_i}^2-(m_{\nu_j}-m_{Z'})^2\right]} \nonumber \\
&\ \ \ \  \times \left\{\frac{(m_{\nu_i}^2-m_{\nu_j}^2)^2}{m_{\nu_i}^2m_{Z'}^2}+1+\frac{m_{\nu_j}^2}{m_{\nu_i}^2}-2\frac{m_{Z'}^2}{m_{\nu_i}^2} \right\}.
\end{align}
Since we assume heavier neutrinos are active ones, the decay rates do not include $g_{ij}^R$ but $g_{ij}^L$. $g_{ij}^R$ appears later in Eq.~(\ref{eq:3bodyboson}) for the decay rate of 3-body decays.
%%%%%%%%%%%%%%%%%%%%%%%%%%%%%%%%%%%%%%%%%%%%%%%%%%%%%%%%%%%%%%%%%%%%%%%%%%%%%%%%%%%%%%%%%%%%%%%%%%
%%%%%%%%%%%%%%%%%%%%%%%%%%%%%%%%%%%%%%%%%%%%%%%%%%%%%%%%%%%%%%%%%%%%%%%%%%%%%%%%%%%%%%%%%%%%%%%%%%
%%%%%%%%%%%%%%%%%%%%%%%%%%%%%%%%%%%%%%%%%%%%%%%%%%%%%%%%%%%%%%%%%%%%%%%%%%%%%%%%%%%%%%%%%%%%%%%%%%
%%%%%%%%%%%%%%%%%%%%%%%%%%%%%%%%%%%%%%%%%%%%%%%%%%%%%%%%%%%%%%%%%%%%%%%%%%%%%%%%%%%%%%%%%%%%%%%%%%

\subsection{3-body decays}
\label{appa2}

\subsubsection{Decays mediated by bosons with spin 0: $\nu_i\rightarrow \nu_j\phi \rightarrow \nu_j\nu_k\bar{\nu}_l$}

The decay rate in each neutrino energy produced by the decays, $d\Gamma_{\nu_i\rightarrow \nu_j\nu_k\bar{\nu}_l}/dE_{\nu_j}$, from the interaction Lagrangian~(\ref{generalphiinteraction}) is 
\begin{align}
\frac{d\Gamma_{\nu_i\rightarrow \nu_j\nu_k\bar{\nu}_l}}{dE_{\nu_j}}&=\frac{2}{\pi^3m_{\phi}^4}\frac{m_{\nu_i}|\bm{p}_{\nu_j}|}{m_{\nu_i}^2-2m_{\nu_i}E_{\nu_j}+m_{\nu_j}^2} \sqrt{(E_{\nu_j}^{\rm max}-E_{\nu_j})\left(E_{\nu_j}^{\rm max}-E_{\nu_j}+\frac{2m_{\nu_k}m_{\nu_l}}{m_{\nu_i}}\right)} \nonumber \\
& \times \biggl\{ (|\lambda_{ij}|^2+|h_{ij}|^2)(|\lambda_{kl}|^2+|h_{kl}|^2)E_{\nu_j}\left[m_{\nu_i}^2+m_{\nu_j}^2-m_{\nu_k}^2-m_{\nu_l}^2-2m_{\nu_i}E_{\nu_j}\right] \nonumber \\
&\ \ \  +2(|\lambda_{ij}|^2+|h_{ij}|^2)(|\lambda_{kl}|^2-|h_{kl}|^2)E_{\nu_j}m_{\nu_k}m_{\nu_l} \nonumber \\
&\ \ \  -(|\lambda_{ij}|^2-|h_{ij}|^2)(|\lambda_{kl}|^2+|h_{kl}|^2)m_{\nu_j}\left[m_{\nu_i}^2+m_{\nu_j}^2-m_{\nu_k}^2-m_{\nu_l}^2-2m_{\nu_i}E_{\nu_j}\right] \nonumber \\
&\ \ \  -2(|\lambda_{ij}|^2-|h_{ij}|^2)(|\lambda_{kl}|^2-|h_{kl}|^2)m_{\nu_j}m_{\nu_k}m_{\nu_l}
\biggl\}. 
\end{align}
Here we assume $j\neq k$, and for $j=k$, we need an additional factor of $1/2$.
We cannot find the analytic formula of the decay rate $\Gamma_{{\nu_i}\rightarrow \nu_j\nu_k\bar{\nu}_l}$ although we can estimate the value of the decay rate numerically as
\begin{align}
\Gamma_{{\nu_i}\rightarrow \nu_j\nu_k\bar{\nu}_l}=\int_{m_{\nu_j}}^{E_{\nu_j}^{\rm max}} dE_{\nu_j} \frac{d\Gamma_{\nu_i\rightarrow \nu_j\nu_k\bar{\nu}_l}}{dE_{\nu_j}}.
\end{align}
For $m_{\nu_i}\gg m_{\nu_j}, m_{\nu_k}, m_{\nu_l}$, we obtain the analytic formula of the decay rate,
\begin{align}
\Gamma_{{\nu_i}\rightarrow \nu_j\nu_k\bar{\nu}_l}\simeq \frac{(|\lambda_{ij}|^2+|h_{ij}|^2)(|\lambda_{kl}|^2+|h_{kl}|^2)}{96\pi^3m_{\phi}^4}m_{\nu_i}^5.
\end{align}

%%%%%%%%%%%%%%%%%%%%%%%%%%%%%%%%%%%%%%%%%%%%%%%%%%%%%%%%%%%%%%%%%%%%%%%%%%%%%%%%%%%%%%%%%%%%%%%%%%
%%%%%%%%%%%%%%%%%%%%%%%%%%%%%%%%%%%%%%%%%%%%%%%%%%%%%%%%%%%%%%%%%%%%%%%%%%%%%%%%%%%%%%%%%%%%%%%%%%

\subsubsection{Decays mediated by bosons with spin 1: $\nu_i\rightarrow \nu_jZ' \rightarrow \nu_j\nu_k\bar{\nu}_l$}

The result of the decay rate in each energy for lighter neutrinos, $d\Gamma_{\nu_i\rightarrow \nu_j\nu_k\bar{\nu}_l}/dE_{\nu_j}$, from the interaction Lagrangian~(\ref{generalZinteraction}) is 
\begin{align}
&\frac{d\Gamma_{\nu_i\rightarrow \nu_j\nu_k\bar{\nu}_l}}{dE_{\nu_j}} \nonumber \\
&=\frac{2}{\pi^3m_{Z'}^4}
\biggl[ |g_{ij}^L|^2|g_{kl}^L|^2\biggl\{\frac{1}{4}(m_{\nu_i}^2+m_{\nu_l}^2-m_{\nu_j}-m_{\nu_k}^2)(E_{\nu_l}^{{\rm max}2}-E_{\nu_l}^{{\rm min}2})-\frac{1}{3}m_{\nu_i}(E_{\nu_l}^{{\rm max}3}-E_{\nu_l}^{{\rm min}3}) \biggl\} \nonumber \\
&\ \ \ \ \ \ \ \ \ \ \ \ +|g_{ij}^L|^2|g_{kl}^R|^2\biggl\{\frac{1}{4}(m_{\nu_i}^2+m_{\nu_k}^2-m_{\nu_j}-m_{\nu_l}^2)(E_{\nu_k}^{{\rm max}2}-E_{\nu_k}^{{\rm min}2})-\frac{1}{3}m_{\nu_i}(E_{\nu_k}^{{\rm max}3}-E_{\nu_k}^{{\rm min}3}) \biggl\} \nonumber \\
&\ \ \ \ \ \ \ \ \ \ \ \ +\frac{1}{2}|g_{ij}^L|^2(g_{kl}^Lg_{kl}^{R\ast}+g_{kl}^{L\ast}g_{kl}^{R})E_{\nu_j}m_{\nu_k}m_{\nu_l}(E_{\nu_l}^{\rm max}-E_{\nu_l}^{\rm min})
\biggl],
\label{eq:3bodyboson}
\end{align}
with $i=1,2,3$. Here we assume $j\neq k$, and for $j=k$, we need an additional factor of $1/2$. $E_{\nu_k}^{\rm max}, E_{\nu_k}^{\rm min}, E_{\nu_l}^{\rm max}$ and $E_{\nu_l}^{\rm min}$ are maximal and minimal energy for $\nu_k$ and $\nu_l$, satisfying the following relations,
\begin{align}
E_{\nu_k}^{\rm max}+E_{\nu_k}^{\rm min}&=\frac{2m_{\nu_i}}{m_{\nu_i}^2-2m_{\nu_i}E_{\nu_j}+m_{\nu_j}^2} (m_{\nu_i}-E_{\nu_j})\biggl[E_{\nu_j}^{\rm max}-E_{\nu_j}+\frac{m_{\nu_k}}{m_{\nu_i}}(m_{\nu_l}+m_{\nu_k}) \biggl],\nonumber \\
E_{\nu_l}^{\rm max}+E_{\nu_l}^{\rm min}&=\frac{2m_{\nu_i}}{m_{\nu_i}^2-2m_{\nu_i}E_{\nu_j}+m_{\nu_j}^2} (m_{\nu_i}-E_{\nu_j})\biggl[E_{\nu_j}^{\rm max}-E_{\nu_j}+\frac{m_{\nu_l}}{m_{\nu_i}}(m_{\nu_l}+m_{\nu_k}) \biggl], \nonumber \\
E_{\nu_k}^{\rm max}-E_{\nu_k}^{\rm min}&=E_{\nu_l}^{\rm max}-E_{\nu_l}^{\rm min} \nonumber \\
&=\frac{2m_{\nu_i} |\bm{p}_{\nu_j}|}{m_{\nu_i}^2-2m_{\nu_i}E_{\nu_j}+m_{\nu_j}^2}(E_{\nu_j}^{\rm max}-E_{\nu_j})^{1/2}\left(E_{\nu_j}^{\rm max}-E_{\nu_j}+\frac{2m_{\nu_k}m_{\nu_l}}{m_{\nu_i}}\right)^{1/2}.
\end{align}
We cannot also find the analytic formula of the decay rate mediated by a gauge boson $Z'$ without additional approximation.
For $m_{\nu_i}\gg m_{\nu_j}, m_{\nu_k}, m_{\nu_l}$, we obtain the analytic formula of the decay rate,
\begin{align}
\Gamma_{\nu_i\rightarrow \nu_j\nu_k\bar{\nu}_l}\simeq\frac{|g_{ij}^L|^2(|g_{kl}^{L}|^2+|g_{kl}^R|^2)}{96\pi^3m_{Z'}^4}m_{\nu_i}^5.
\end{align}

%%%%%%%%%%%%%%%%%%%%%%%%%%%%%%%%%%%%%%%%%%%%%%%%%%%%%%%%%%%%%%%%%%%%%%%%%%%%%%%%%%%%%%%%%%%%%%%%%%
%%%%%%%%%%%%%%%%%%%%%%%%%%%%%%%%%%%%%%%%%%%%%%%%%%%%%%%%%%%%%%%%%%%%%%%%%%%%%%%%%%%%%%%%%%%%%%%%%%
%%%%%%%%%%%%%%%%%%%%%%%%%%%%%%%%%%%%%%%%%%%%%%%%%%%%%%%%%%%%%%%%%%%%%%%%%%%%%%%%%%%%%%%%%%%%%%%%%%
%%%%%%%%%%%%%%%%%%%%%%%%%%%%%%%%%%%%%%%%%%%%%%%%%%%%%%%%%%%%%%%%%%%%%%%%%%%%%%%%%%%%%%%%%%%%%%%%%%

\end{document}